\documentclass[twocolumn]{aastex63}

\usepackage{graphicx}
\usepackage{subfigure}

\usepackage{hyperref}
\usepackage{natbib}
\usepackage{amsmath}
\usepackage{mathrsfs}

\usepackage{etoolbox}

\AtBeginEnvironment{quote}{\singlespace\vspace{-\topsep}\small}
\AtEndEnvironment{quote}{\vspace{-\topsep}\endsinglespace}

\let\oldAA\AA
\renewcommand{\AA}{\text{\normalfont\oldAA}}

\begin{document}
\title{The Assembly History of M87 Through Radial Variations in Chemical Abundances of its Field Star And Globular Cluster Populations}

\author[0000-0003-1887-0621]{Alexa Villaume}
\affiliation{Department of Astronomy \& Astrophysics, University of California Santa Cruz, 1156 High Street, Santa Cruz, CA 95064, USA}
\correspondingauthor{Alexa Villaume}
\email{avillaum@ucsc.edu}

\author[0000-0002-9328-5652]{Daniel Foreman-Mackey}
\affiliation{ Center for Computational Astrophysics, Flatiron Institute, 162 5th Ave, New York, NY 10010, USA}

\author[0000-0003-2473-0369]{Aaron J. Romanowsky}
\affiliation{Department of Physics \& Astronomy, San Jose State University, One Washington Square, San Jose, CA 95192, USA} 
\affiliation{University of California Observatories, 1156 High Street, Santa Cruz, CA, 95064, CA, USA}

\author[0000-0002-9658-8763]{Jean Brodie}
\affiliation{Department of Astronomy \& Astrophysics, University of California Santa Cruz, 1156 High Street, Santa Cruz, CA 95064, USA}

\author[0000-0002-1468-9668]{Jay Strader}
\affiliation{Center for Data Intensive and Time Domain Astronomy,
Department of Physics and Astronomy,
Michigan State University,
East Lansing, MI 48824, USA}

\begin{abstract}

We present an extensive study of spectroscopically-derived  chemical abundances for M87 and its globular cluster (GC) system. Using observations from the Mitchell spectrograph at McDonald, LRIS at Keck, and Hectospec on the MMT, we derive new metallicity gradients from $\sim 2$ to $140$ kpc. We use a novel  hierarchical statistical framework to simultaneously separate the GC system into subpopulations while measuring the metallicity gradients of those subpopulations. We create physically-motivated spectral stacks of the GC subpopulations by leveraging the output of this statistical framework to perform the first application of abundance tagging in a massive ETG to better constrain the origins of the GC subpopulations and, thus, the assembly history of M87. We find a metal-poor, $\alpha$-enhanced population of GCs in both in the inner and outer halo unanticipated by current cosmological simulations of galaxy evolution. We use the remarkably flat metallicity gradients we find for both the metal-rich and metal-poor GC subpopulations in the inner halo as tentative evidence that some amount of the metal-poor GCs formed directly in the halo of M87 at high redshift.
\end{abstract}


\section{Introduction}

When it comes to emphasizing the importance of understanding galaxy evolution, and the difficulties therein, it cannot get much better than \citet{tinsley1980} review: ``{\it Essentially everything of astronomical interest is either part of a galaxy, or from a galaxy, or otherwise relevant to the origin and evolution of galaxies...This is not a field in which one can hope to develop a complete theory from a simple set of assumptions, because many relevant data are unavailable or ambiguous, and galactic evolution depends on many complicated dynamical, atomic, and nuclear processes which themselves are incompletely understood.}''

\noindent In the subsequent 40 years, many significant advances in observations, computation, and theory have been made but the fundamental problem described by \citet{tinsley1980} remains. The huge range in all manner of relevant physical scales from time, to size, to mass, make it impossible to establish an {\it a priori} model of galaxy evolution \citep[for a modern take see][and the references therein]{somerville2015, naab2017}. Which is not to say that no progress has been made. $\Lambda$CDM cosmology is now the conventional paradigm and so galaxy evolution is now viewed to be fundamentally hierarchical, that after an initial phase of {\it in-situ} star formation, an extended second phase of accretion of lower mass satellites bring in an {\it ex-situ} population of stars which build the stellar halos in more massive galaxies. 

The expectation then is that the stellar halos of galaxies contain a wealth of information about their history. The ``archaelogical'' approach is use the chemistry and dynamics of long-lived stars to uncover this history \citep[e.g.,][and the references therein]{helmi2020}.
In recent years, our understanding of the origins of the Milky Way has been transformed through a combination of heroic data collection efforts \citep[e.g. the Gaia, APOGEE surveys;][]{gaia2018, majewski2017} and increasingly sophisticated simulations of stellar halos in a  cosmological context \citep[e.g., Latte/FIRE2, Auriga;][]{wetzel2016, grand2017}.
But it is massive early-type galaxies (ETGs) that undergo the most active satellite infall and therefore provide a key constraint for the theoretical understanding of this process \citep{delucia2007}.

In light of this, there has been a push to obtain galactocentric radial gradients of stellar population parameters. Despite the long history of using gradients to falsify galaxy formation scenarios \citep[see an early review by][]{faber1977}, this field is only recently reaching maturity due to advances in integral field unit (IFU) spectrographs \citep[e.g., the MASSIVE and MaNGA surveys;][]{ma2014, bundy2015}.
Even with these advances, gradients for massive ETGs still only extend out to just a couple effective radii with the galaxy light \citep[e.g.,][]{greene2013}.

It is difficult to fully constrain the characteristics of the {\it ex-situ} population in this region because of the possible contamination of an {\it in-situ} population \citep[see discussion in][]{greene2019}.
Unlike the Milky Way, where individual resolved stars are accessible, beyond Local Group only integrated light of the unresolved stellar populations is accessible spectroscopically.
Imaging surveys can now reach the  ``outer halos'' of massive ETGs \citep[e.g, the Burrell-Schmidt Survey and Hyper Suprime-Cam surveys;][]{mihos2017, aihara2018}, which are expected be dominated by {\it ex-situ} stars. However, without detailed chemistry and dynamics of the stellar population, not much progress can be made to quantify the assembly histories of these galaxies.

It will not be until the next generation of telescopes come online will it be possible to obtain spectroscopy with enough signal to measure measure detailed stellar population properties in the outer halo. In the meantime, instead of focusing solely on the stellar populations within a galaxy, globular clusters (GCs) can be used as ``discrete tracers''. GCs are nearly ubiquitous around galaxies and are relatively luminous compared to galaxy starlight.
But their strength lies in their uniformly old ($>10$ Gyr) ages which makes them an ideal ``fossil record'' for the archaeological approach as they presumably reflect the early conditions under which they formed.

In the Milky Way, the GC system provided the pre-$\Lambda$CDM evidence that the stellar halo was built through the accretion of satellite galaxies \citep{searle1978}. Moreover, the difference in accretion histories between the Milky Way and M31 is clearly seen in the different metallicity distributions of their respective GC systems \citep[][]{caldwell2016}. GC systems also extend far beyond the reaches of galaxy starlight, and so provide a window to the fully {\it ex-situ} outer halos of galaxies while at the same time providing an independent probe of the more complicated inner halos.

The brightest cluster galaxy (BCG) of the Virgo Cluster, M87 (NGC 4486), has one of the most extensively studied GC systems \citep[starting with][]{baum1955}, both photometrically \citep[e.g.,][]{peng2006, strader2011b} and kinematically \citep[e.g.,][]{romanowsky2012, oldham2016} but not archaeologically. This is reflective of broader landscape of our understanding extragalactic GC systems which has been done primarily with broadband photometry. The  most extensive {\it spectroscopic} work to date has been through the SLUGGS  Survey \citep{brodie2014} which focused on kinematics.

Similarly, the stellar populations of the galaxy light of M87 itself have remarkably never been studied with spectroscopy beyond the central few kpc. In this work, we jointly analyze the spatially-resolved stellar population properties M87 out to $\sim 20$ kpc using archival IFU data and its GC system out to $\sim 140$ kpc using a fully spectroscopic sample. We present a statistical framework to characterize the GC system as an aggregate of subpopulations to achieve more accurate inferences of the physical parameters of the GC subpopulations, particularly the metallicity gradients, than previous studies.

We take the distance to M87 to be $D_L = 16.5$ Mpc, with effective radius $R_e = 16.0$ kpc \citep{kormendy2009}, and ${\rm log} (M_*/M_\odot ) = 11.61 \pm 0.10$ \citep{oldham2016b}. In Section 2, we describe the spectroscopic samples and the stellar population synthesis models we use to extract abundance information from both M87 and its GC system. In Section 3, we  motivate the need for a new approach to measure metallicity gradients in GC subpopulations and outline a novel statistical framework to make this measurement. In Section 4, we present the measured metallicity gradients for the GC subpopulations, the detailed stellar population gradients of the M87 starlight, detailed chemistry of the GC population using stacked spectra. In Section 5, we discuss the results to understand the progenitor populations of the stellar halo and some aspects of the origins of the metals in M87 and, finally, in Section 6, we summarize our results and highlight our main conclusions.

\section{Spectroscopic Data and Abundance Analysis}

\subsection{Obtaining the stellar population parameters} 

We model the spectra with an updated version of the absorption line fitter \citep[\texttt{alf,}][]{conroy2012, choi2014, conroy2014, conroy2018a}\footnote{https://github.com/cconroy20/alf} that uses the Extended IRTF Library \citep[E-IRTF;][]{villaume2017a} and the MIST isochrones \citep[][]{choi2016}. With \texttt{alf} we can model the full continuum-normalized spectrum of integrated light for stellar ages $> 1$Gyr and for metallicities $\sim -2.0$ to $+0.25$. The full model has 36 free parameters \citep[see Table 2 in ][]{conroy2018a}. The parameter space is explored using a Markov Chain Monte Carlo algorithm \citep[\texttt{emcee;}][]{emcee_v1}. In this work we use the priors as described in \citet{conroy2018a} and fix the IMF to the  \citet{kroupa2001} form. 

Theoretical elemental response functions  that tabulate the effect on the spectrum of enhancing each individual element modeled in \texttt{alf} were computed with the ATLAS and SYNTHE programs \citep{atlas, synthe}. For the $\alpha$ elements relative to Fe considered in our analysis (Mg, Si, and Ca) we correct for the underlying abundance pattern in the empirical stellar library using the [Mg/Fe] values from \citet{milone2011} and [Ca/Fe] values from \citet{bensby2014}. We assume Si has the same library abundance pattern as Ca. 

We analyze several different data sets in this work (see below). To make the different samples as homogeneous as possible we fitted over the same spectral range for every spectrum analyzed in this work: $4000 < \lambda \AA < 4400$ and  $4400 < \lambda \AA < 5225$. 

While we obtain estimates of the light-weighted age as part of the \texttt{alf} models, we do not include age in our analysis. This is because of the uncertain effect of the blue horizontal branch, particularly in the GCs, which can make the inferred ages artificially young. Our analysis of the Milky Way GC system indicates that iron metallicity can still be reliably recovered in the presence of a blue horizontal branch \citep[see][]{conroy2018a}.

\subsection{The globular clusters}

\begin{figure}
\includegraphics[width=0.5\textwidth]{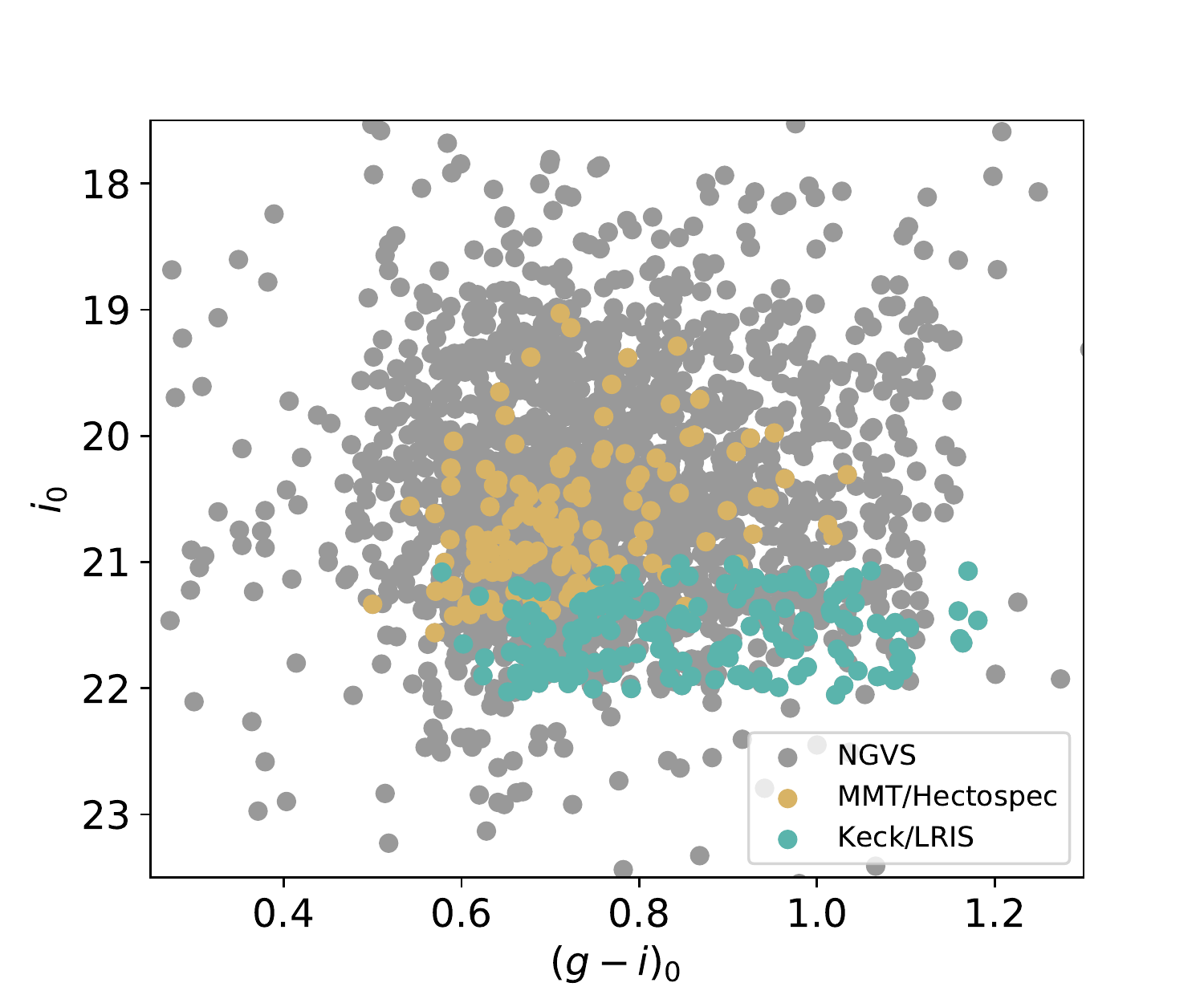}
\caption{Comparing the coverage of the Keck/LRIS (green)  and MMT/Hectospec (yellow) samples in color-magnitude space. Also shown is the NGVS sample \citep[][grey]{oldham2016a}. The MMT/Hectospec sample is overall more luminous than the Keck/LRIS sample and has more blue GCs than red, while the Keck/LRIS sample is evenly distributed over color space.
}
\label{fig:cmd}
\end{figure}

\citet{strader2011b} carried out a wide-field kinematic analysis of the M87 GC system using two key data sets: the Keck/LRIS sample and the MMT/Hectospec sample.  In Figure~\ref{fig:cmd} we compare the two samples in color--magnitude space. The Keck/LRIS sample (green) was selected to sample the low luminosity population over the full color range of the GC system, in contrast to previous work that targeted high-luminosity clusters that likely have different properties from the bulk of the GCs \citep[see discussion in][]{villaume2019}. The MMT/Hectospec sample (yellow) was selected from the higher luminosity population. The MMT/Hectospec objects were also primarily selected at large radii to aid the sky subtraction since Hectospec is a fiber instrument.

In \citet{villaume2019} we applied full-spectrum stellar population synthesis (SPS) models to the Keck/LRIS dataset of M87 GCs \citep{strader2011b} to obtain estimates of iron metallicity ([Fe/H]) relative to solar. We refer the readers to the original paper for details on the modeling and validation of the [Fe/H] values for the Keck/LRIS sample. Here, we do the same analysis for the MMT/Hectospec sample. We used the square root of the summed sky spectrum and flux generated by the reduction pipeline as the uncertainty on the individual GCs. The S/N of this data set ranges from S/N $\sim 1-30 {\AA^{-1}}$ with a resolution of $5{\AA}$. The  resolution of the data is higher than the native resolution of the models so we smoothed the data to 200 km/s to be consistent with our previous analysis with the Keck/LRIS data. 

\begin{figure*}
\includegraphics[width=1\textwidth]{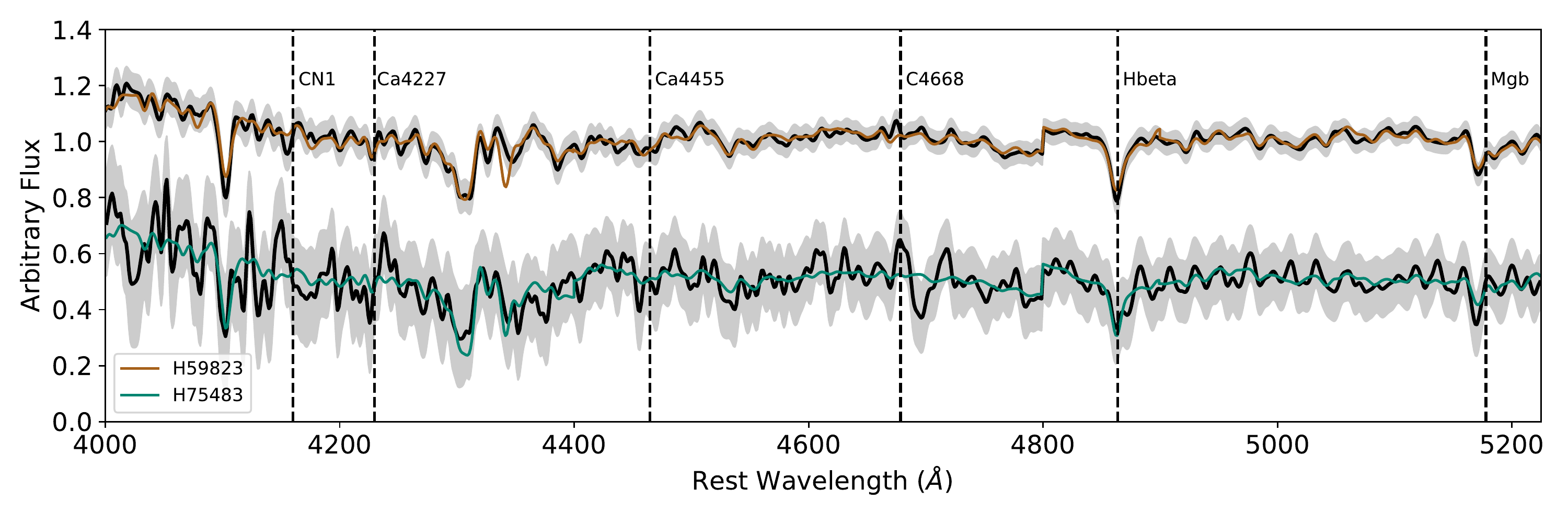}
\caption{Comparison of MMT/Hectospec GC spectra (black) and best-fit models for a comparatively high-S/N spectrum (S/N$\sim 30$, brown) and a low-S/N spectrum (S/N$\sim10$, green). Grey band is the uncertainty of the flux from the input spectrum. Within the uncertainties, the fits are successful. 
}
\label{fig:mmt_spec}
\end{figure*}

Before we smoothed, we identified particularly bad sky lines in the spectra at $4040 < \lambda {\rm \AA} < 4050$, $4355 < \lambda  {\rm \AA} < 4365$, and $5458 < \lambda  {\rm \AA} < 5470$ and interpolated over the flux in each spectrum in those wavelength regions. We fitted 156 spectra and rejected 12 spectra from our analysis based on visual inspection of the residuals between the observed spectra and best-fit models. We show successful fits in Figure~\ref{fig:mmt_spec} for a comparatively high-S/N spectrum (S/N$\sim 30$, brown) and a low-S/N spectrum (S/N$\sim10$, green) with spectral features of particular interest highlighted. The black line and grey band represent the data flux and uncertainty, respectively.

For the individual GCs, we focus our analysis on [Fe/H] and summarize our measurements in Table~\ref{table:feh_measurements}. The majority of the GC spectra do not have sufficient S/N to reliably extract more detailed abundance information. In Section 4.3 we describe how we stacked the individual GC spectra and fit the stacks with \texttt{alf}. 

\begin{deluxetable}{cccccc}
\tabletypesize{\footnotesize}
\tablecolumns{8}
\tablewidth{0.9\columnwidth}
\tablecaption{Table of summary statistics of the [Fe/H] measurements for the GCs included in this work$^\dagger$. \label{table:feh_measurements}}
\tablehead{
 \colhead{ID}  & \colhead{RA} & \colhead{DEC} & \colhead{[Fe/H]} & \colhead{$\sigma_{\rm [Fe/H]}$ } & \colhead{Instrument} \\
}
\startdata
\centering
H47487 & 187.73553 & 12.32802 & $-$1.16 & 0.30 & LRIS \\
H49585 & 187.67674 & 12.32961 & $-$0.51 & 0.41 & LRIS \\
H49328 & 187.71423 & 12.32992 & $-$0.78 & 0.24 & LRIS \\
... & & & \\
H47487 & 187.59446 & 12.02249 & $-$0.80 & 0.37 & Hectospec \\
H49585 & 187.52539 & 12.03362 & $-$0.40 & 0.23 & Hectospec \\
H49328 & 187.91104 & 12.04288 & $-$1.17 & 0.39 & Hectospec \\
\enddata
\tablecomments{This table is available in its entirety in a machine-readable form in the online journal. A portion is shown here for guidance regarding its form and content.}
\tablenotetext{\dagger} {Full [Fe/H] posteriors of all GC measurements can be found: \url{github.com/AlexaVillaume/m87-gc-feh-posteriors}}
\end{deluxetable}

\subsection{The galaxy light}

\begin{figure*}
\includegraphics[width=1\textwidth]{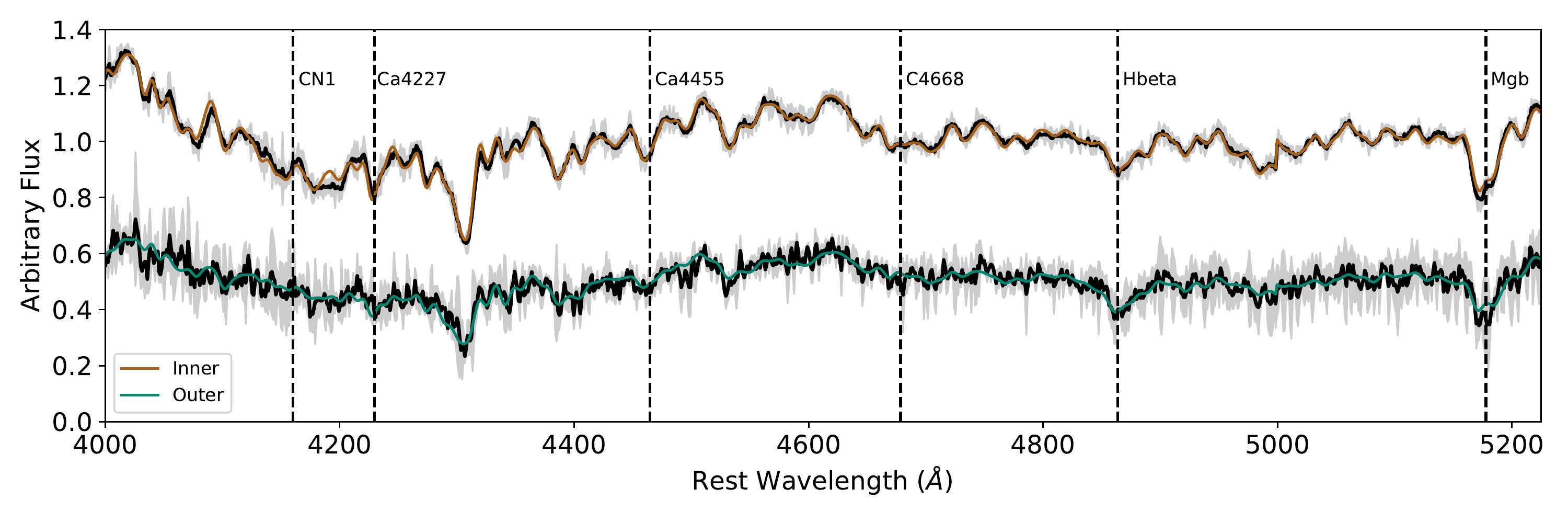}
\caption{Same as Figure~\ref{fig:mmt_spec} but for Mitchell spectra close to the center of the galaxy (R$_{\rm gal} =  1.32$ kpc, brown) and from the outer region (R$_{\rm gal} = 19.5$ kpc, green).
}
\label{fig:vp_spec}
\end{figure*}

We use data from the Mitchell (formerly VIRUS-P) integral field unit (IFU) spectrograph at McDonald Observatory (\citealt{murphy2011}; spectroscopy obtained via private communication with K.~Gebhardt). The signal-to-noise (S/N) of the individual spectra ranges from $\sim 20 - 50 {\rm \AA}^{-1} $. We stacked the individual spectra in 10 bins of galactocentric radius by bootstrapping for the median of the individual spectra in a given bin. We used the 50th percentile from the resulting distribution of flux at a given wavelength as the stacked spectrum and used the average of the 16th and 84th percentiles as the uncertainties on the stacks with the S/N of the stacks ranging from $\sim 40-200$, with the outermost spectrum having the lowest S/N.

 \begin{figure*}
\includegraphics[width=1\textwidth]{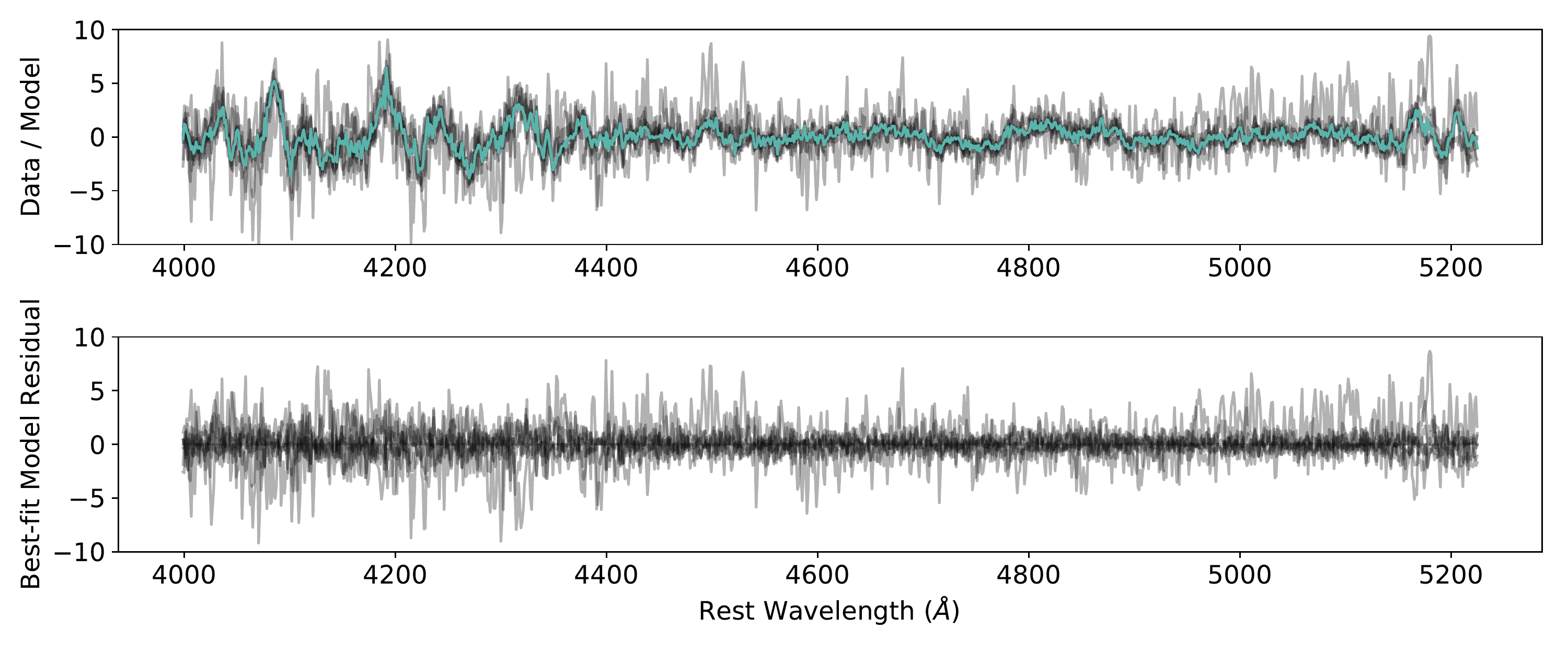}
\caption{(Top) Residuals from dividing the best-fit models from the corresponding  Mitchell integrated galaxy light (black) and the median residual for all spectra (green). The residuals are nearly identical for all spectra and the large wavelength-scale features are likely systematic to the models  and not dependent on stellar parameters. (Bottom) Residuals after subtracting the median residual.}
\label{fig:vp_residuals}
\end{figure*}

In Figure~\ref{fig:vp_spec} we examine the quality of our fits for spectra in the inner (R$_{\rm gal} \sim 1.32$ kpc, S/N$\sim 200$) part of the galaxy and the outer part (R$_{\rm gal} \sim 19.4$ kpc, S/N$\sim 40$). We compare the Mitchell spectra (black) with the best-fit model spectrum for the inner region (brown) and the outer region spectrum (green) with selected spectral features highlighted. The grey bands are the flux uncertainty from the data. In Figure~\ref{fig:vp_residuals} we examine the residuals between the best-fit model and input data for all the Mitchell spectra used in this analysis. The residuals are typically small ($< 5\%$).

\section{Characterizing Globular Cluster Systems Via Statistical Modeling}
\label{sec:model}
Our goal is to develop a method to measure the properties of GC systems as a way to understand the formation history of M87 and other galaxies. In this work, we are focused on measuring the metallicity gradients and abundance patterns of the M87 GC system.
Obtaining a metallicity gradient might seem as simple as fitting a line to data, but a recent meta-analysis of many of the studies that have measured metallicity gradients of GC systems revealed a troubling result -- different studies often get significantly different answers for the {\it same} GC systems \citep[see Figure 1 in][]{forbes2018}. Several underlying issues could be causing an accuracy problem in these studies, which motivates us to characterize GC systems in a novel way using a hierarchical Bayesian model (HBM). In the following, we detail these issues and describe how HBMs provide a natural means to overcome them. 

First, the studies included in the \citet{forbes2018} analysis all used a version of linear least-squares to fit the gradients of GC studies. However, linear least-squares only works if one of the dimensions of data has negligible uncertainties. These studies also assumed that the galactocentric distances of the GCs are perfectly known. This is not the case, however, since only 2D {\it projected} distances are known. The distances can be de-projected if the 3D density distributions of the GCs are known \citep[e.g.][]{mclaughlin1999}, but this is not the case for the vast majority of extragalactic GC observations. As discussed in \citet{liu2011}, using the projected distances as a substitute for true distance introduces systematic uncertainty into the measured gradients because the GCs projected into the center will, in reality, be a mix of GCs at all radii. \citet{liu2011} estimated that this could lead to an uncertainty of $\sim 10\%$ in the measured gradients, but in reality, this depends on the degree of the true underlying slope. We must take that uncertainty into account when interpreting the measured gradients.

Second, characterizing GC systems is further complicated because these systems, especially around massive galaxies, are the aggregate of many different stellar populations. The constraints on GC system assembly and galaxy formation depends on our ability to differentiate and understand the {\it subpopulations} of a GC system. Broadly, GCs are separable into ``metal-poor'' and ``metal-rich'' populations. In detail, however, it is not trivial to separate the individual GCs into subpopulations.

Previous work measuring the metallicity gradients of GC systems has primarily used constant cuts on color to separate the metal-poor and metal-rich subpopulations \citep[e.g.,][]{harris2009a, harris2009b, liu2011, hargis2014, kartha2016}. However, wide-field photometric surveys have demonstrated that the demographics of GC populations change with increasing distance from the center of the galaxy \citep[e.g.][]{strader2011b, harris2017a},  with the relative number of blue GCs typically increasing. As a result, a constant cut across the GC sample could bias the gradient measurements (see later in this section for demonstration of this effect). A few studies have attempted to mitigate this issue by separating the GC subpopulations at different radial steps \citep[e.g.,][]{blom2012, usher2013}. However, these studies did not measure the gradient for their full samples but only considered the peaks of the metallicity distribution functions (MDFs) when computing the gradients. Moreover, by cutting on subpopulation {\it and then} determining subpopulation characteristics, all these studies fail to account for the covariance between subpopulation membership assignments and whatever parameter of interest is being measured. This, again, will bias the gradient measurements.

Finally, linear least-squares is highly sensitive to the presence of outliers in a sample. The studies included in the \citet{forbes2018} analysis used photometric samples of GCs with colors as a proxy for stellar metallicity, except for \citet{pastorello2015} who also had calcium triplet (CaT) determined metallicities. Without spectroscopic follow-up to confirm GC candidates in photometric surveys, any study based on such data will be affected by contaminant populations. Furthermore, the color--metallicity relations that are used to convert broadband colors of GCs into iron ([Fe/H]) metallicities have been recently called into question \citep{usher2012, villaume2019}. In this work, we use only spectroscopically-determined [Fe/H] measurements of the individual GCs.

HBM provides a means to address and mitigate these issues.
Specifically, HBM is a natural way to fit the galactocentric metallicity gradients of GC systems for a number of reasons:
\begin{itemize}
\item The Bayesian framework allows us to model unobserved (latent) parameters. This means we can directly model and fit any intrinsic scatter in the metallicities as an explicit parameter and {\it marginalize} over the unknown 3D distribution of the GC system to mitigate the bias from the projected distances.
\item We do not need to make {\it a priori} cuts to obtain the subpopulations. Instead, we can fit the linear metallicity gradients jointly with the subpopulation memberships, allowing us to capture the covariance between the subpopulation slopes and the subpopulation memberships. This helps us obtain more accurate subpopulation membership assignments for the individual GCs and, thus, more accurate metallicity gradients.
\item Relatedly, instead of making a binary cut with the subpopulation assignments, we get probabilistic memberships. We can propagate the uncertainties on the subpopulation membership assignments throughout this work. This is especially important because we are also interested in the detailed abundance patterns of the GC subpopulations. Currently, the signal-to-noise (S/N) of the spectroscopy does not allow for reliable estimates of abundances for the majority of the individual GCs in our sample, so creating spectral stacks with reliable uncertainties is crucial.
\item Moreover, with HBM, like all Bayesian methods, we produce posterior distributions for {\it all} the model parameters. In practice, this gives us trustworthy and interpretable uncertainties on the gradient measurements.
\end{itemize}

\noindent In short, HBM provides us with results that are more accurate and interpretable, with uncertainties which better represent the reality than previous studies. In the rest of this section, we develop a method that allows this full propagation of uncertainty from the measurements to the inferences made about the subpopulation distributions and demonstrate its efficacy with mock data.

In the following Section we provide a pedagogical explanation of our model as a way to introduce HBM. For those already familiar with this statistical technique, our full model is collectively summarized in Table~\ref{table:priors}, Figure~\ref{fig:multipop_viz}, and Equation 7.

\begin{figure}
\includegraphics[width=0.5\textwidth]{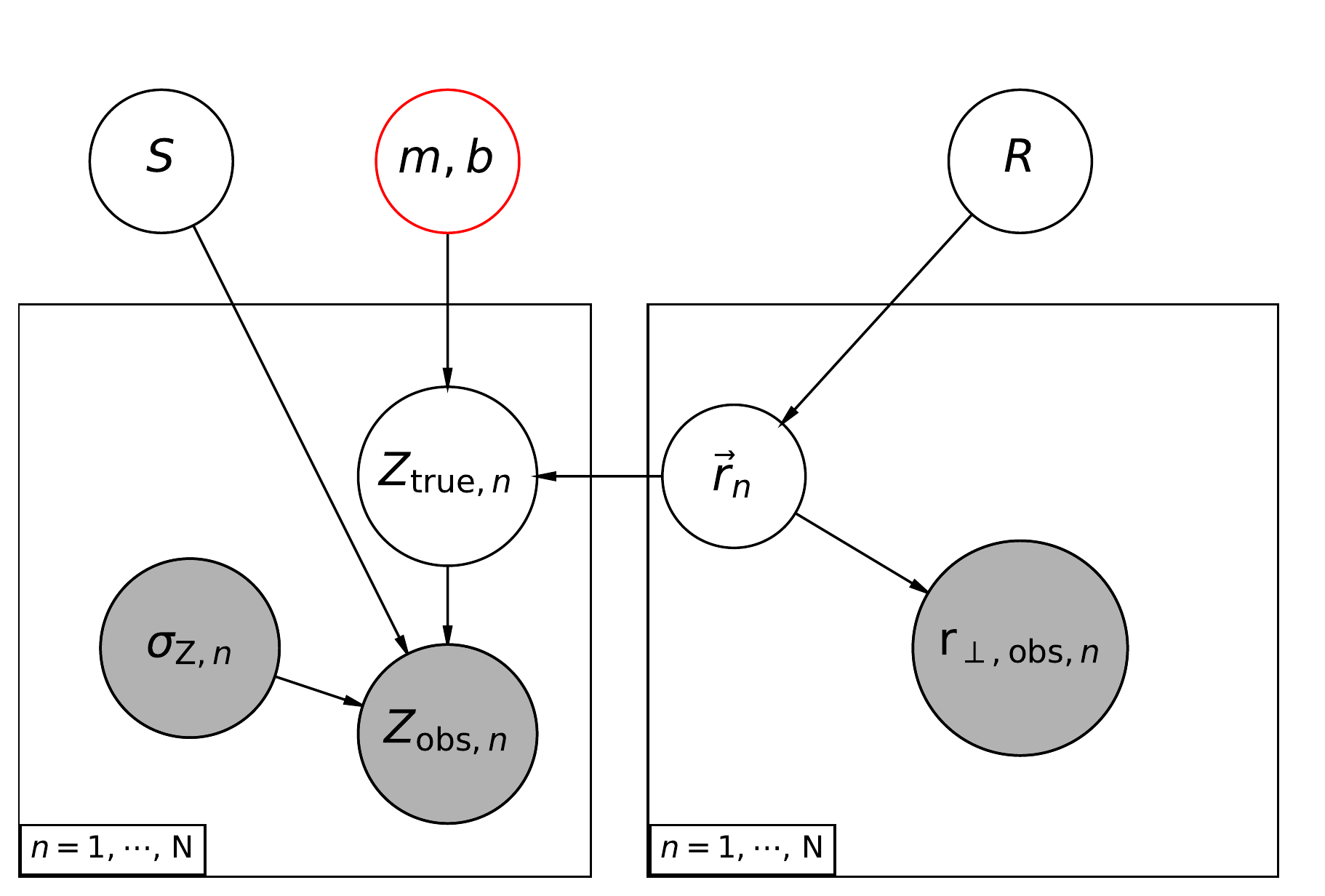}
\caption{The graphical representation of our single population model that we use to factorize the joint distribution of our model. We condition on the observations (grey) to make inferences about the latent parameters (open circles) parameters of interest, the slope, $m$, and intercept, $b$ (red circle).
The rectangle (``plate'') represents the structure of the individual parameters and data that is repeated for all of the GCs in our sample ($n=1,...,N$). The arrows show the direction of conditional dependence among the parameters. See Section 3.1 for details on the parameters.
}
\label{fig:singlepop_viz}
\end{figure}

\begin{deluxetable*}{lll}
\tabletypesize{\footnotesize}
\tablecolumns{8}
\tablewidth{0.9\columnwidth}
\tablecaption{Table of parameters for the full hierarchical mixture model with their prior distributions and qualitative description of their purpose in the model. \label{table:priors}}
\tablehead{
 \colhead{Probability density distributions}  & \colhead{Prior} & \colhead{Description}
}
\startdata
\centering
$p(Z_{{\rm obs, n}} | Z_{{\rm true, n}}, S, \sigma_n)$  & Normal($Z_{{\rm true, n}}$, $\sqrt{S^2 + \sigma_n^2}$) & Observations are a noisy realization of true values \\
$p(\sigma_n)$  & Delta$(\sigma_n)$  & Measured uncertainty on measurements \\
$p(S_c)$ & log Normal($-1.0$, $1.0$) & Unmeasured uncertainty in gradient\\
$p(Z_{{\rm true}, n} | m,  b, \vec{r}_n)$ & Deterministic($m \times |\vec{r}_n| +  b$) & True values come from an underlying gradient\\
$p(m_c, b_c)$ & Deterministic(tan$^{-1}$($\theta)$, $b_\perp/{\rm cos}(\theta)$) & The covariant gradient parameters\\
$p(\theta)^\dagger$ & Uniform($-0.5\pi$, $+0.5\pi$) & Angle between gradient and horizontal axis\\
$p(b_\perp)^\dagger$ & Uniform($-10$, $10$) & Perpendicular intercept \\
$p(r_{\perp, obs, n} | \vec{r})$ & Delta$(r_{\perp, {\rm obs}} - \sqrt{r_x^2 + r_y^2})$ & Observed distances are a realization of true 3D distance \\
$p(\vec{r_n} | R_c)$ & Isotropic Normal & True 3D distance \\
$p(R_c)$ & log Normal($10$, $5$) & Scale length of the subpopulation \\
$q_n$ & Categorical($P_c$) & Subpopulation membership identifier\\
\enddata
\tablenotetext{\dagger} { Uniform priors in $m$ can potentially bias the results toward higher slopes, to avoid this we fit the gradient using the angle between the line $\theta = {\rm Tan}^{-1}(m)$, and the ``perpendicular intercept'' $b=b_\perp/{\rm Cos}(\theta)$. \\ \url{http://jakevdp.github.io/blog/2014/06/14/frequentism-and-bayesianism-4-bayesian-in-python/}}
\end{deluxetable*}

\subsection{A model for a single population}
\label{sec:singlepop}

We begin with a model for a single population of objects as a way to demonstrate some of the key reasons for using a HBM framework in a simplified setting. Bayesian inference is an application of Bayes' theorem \citep{bayes1763},  

\begin{equation}
    P(A|B) = \frac{P(B|A)P(A)}{P(B)}\text{, if } P(B) \neq 0. 
\end{equation}

\noindent which is derived from an axiom of conditional probability. Bayes' Theorem is just a way to compute conditional probabilities of events while folding in prior knowledge related to that event. In practice as a tool for statistical inference, Bayes' theorem is often written in terms of parameters, $\theta$, and data, $x$, and the denominator, also known as the Bayesian evidence is often dropped to yield the {\it unnormalized posterior density}, $p(\theta|x) \propto p(x|\theta)p(\theta)$ \citep{gelman_bda}.

The first term on right-hand side of the proportionality is the {\it likelihood function} and the second term is the {\it prior distribution}. The likelihood function describes the connection between the available data and the parameters of interest.

In this work, the data we have is $x = [r_{\perp, {\rm obs}, n}, Z_{{\rm obs}, n}, \sigma_{{\rm Z}, n}]$ for each $n$ GC and our ultimate parameters of interest are the slope, $m$, and intercept $b$ (highlighted with a red node and are together in the same node to indicate their covariance) of the metallicity gradient. However, we construct our model based on the idea that the data we have correspond to true versions of the parameters that are obfuscated by systematic and random uncertainty. This introduces {\it latent}, i.e., unobserved, parameters to our model. That is, the observed metallicity of a GC, $Z_{{\rm obs, n}}$, is a noisy realization of that GC's true metallicity, $Z_{{\rm true, n}}$ and its observed projected distance, $r_{\perp, {\rm obs}, n}$, is a realization of the true 3D distance, $|\vec{r}_n|$.

In the left-hand side of  Figure~\ref{fig:singlepop_viz}, we show part of the graphical representation of our probabilistic model ($|\vec{r}_n|$  will be discussed in more detail later). This shows how the relationship between the observations (filled nodes) and the parameters relevant to the inference we want to make (open nodes). Within the rectangle (known as the ``plate''), we show the data and parameters for the individual GCs in the sample. The parameters outside the plate are the parameters for the whole population. We will now distinguish these population parameters (the {\it hyperparameters}, $\alpha$) from the parameters for the individual GCs ($\theta_n$). With the introduction of the $\alpha$ parameters, the joint distribution we seek to constrain is $p(\alpha, \theta_n | x_n)$, to which we can apply Bayes' Theorem:,

\begin{equation}
    p(\alpha, \theta_n | x_n) \propto p(x_n | \theta_n, \alpha) p(\theta_n, \alpha).
\end{equation}

In Figure~\ref{fig:singlepop_viz} the arrows represent the conditional dependency among the different parameters and makes clear the hierarchical nature of our model. The key point is that the data, $x_n$, are only conditionally dependent on the parameters $\theta_n$ and are therefore {\it conditionally independent} from the hyperparameters, $\alpha$. That, and being able to factor $p(\theta_n, \alpha)$ to $p(\theta_n|\alpha)p(\alpha)$ gives,

\begin{equation}
     p(\alpha, \theta_n | x_n) \propto p(x_n | \theta_n) p (\theta_n | \alpha) p(\alpha).
\end{equation}

\noindent The key difference between standard Bayesian models and HBMs is that we constrain the population parameters by conditioning on the observations of the many individual GCs rather than fixing them and using them as priors.

The gradient parameters are inferred through modeling the ``true'' metallicity values of the individual GCs, $Z_{{\rm true}, n}$.
The $Z_{{\rm true}, n}$ values are set deterministically by the linear relation  p($Z_{{\rm true}, n} | m, b, \vec{r_n}) = m \times |\vec{r}_n| +  b$.
We condition  $Z_{{\rm true}, n}$ on $Z_{{\rm obs}, n}$ by modeling the observations as drawn from normals centered on the true values and with a standard deviation that encompasses our uncertainty on the metallicity gradient. This uncertainty is the quadrature sum of the uncertainties on the individual [Fe/H] measurements, $\sigma_{Z, n}$, and {\it unobserved} uncertainty for the intrinsic scatter in the radial metallicity gradient, $S$, such that $\sigma = \sqrt{S^2 + \sigma_{Z, n}^2}$.

With the $\vec{r}_n$ dependence for $Z_{{\rm true}, n}$ we introduce a key advantage when using a Bayesian framework. Even though we do not have the line-of-sight distances, $r_{\parallel, n}$, we can make inferences on the true distances for each GC while only making weak assumptions about the population.
Specifically, we model the angular distribution of GCs as isotropic and assume that the GCs are normally distributed in radius by some scale length, $R$, in all 3 coordinates $xyz$ and marginalize over the angle, $\phi_n$, between $x_n$ and $y_n$ to get,  
\begin{multline}
p(\vec{r}_n | R) = \frac{r_{\perp, n}}{R^2}{\rm exp}\left[\frac{-r_{\perp, n}^2}{2R^2}\right] \times \\ \frac{1}{\sqrt{2\pi R^2}}{\rm exp}\left[\frac{-r_{\parallel, n}^2}{2R^2}\right],
\label{eq:dlike}
\end{multline}

\noindent which we fully derive in Appendix~\ref{ap:derivation}. As such, we model the projected distances as drawn from a Rayleigh distribution (the first term on the right-hand side of the above equation) and the line-of-sight distances as drawn from a normal distribution (the second term). This structure is graphically represented in the right-hand side of Figure~\ref{fig:singlepop_viz}.

\begin{figure*}
\includegraphics[width=1\textwidth]{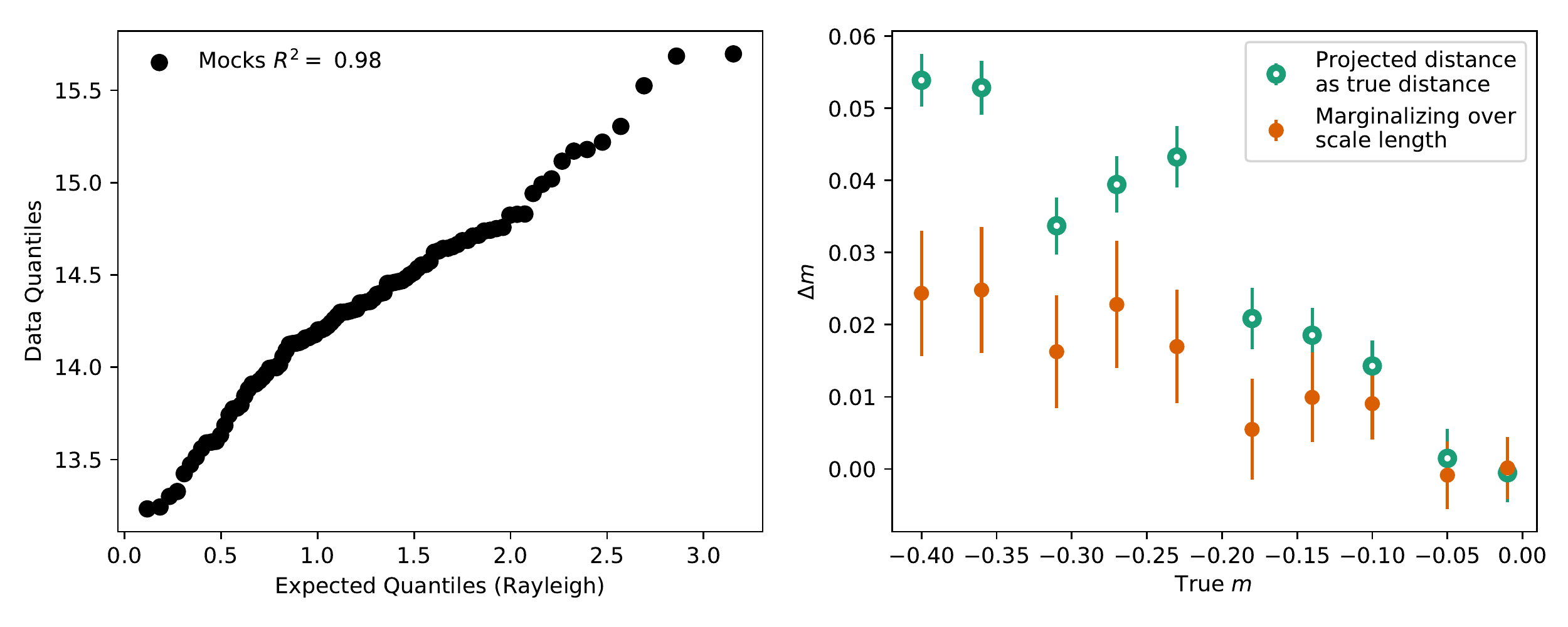}
\caption{(Left) Expected quantiles of a Rayleigh distribution versus quantiles of the projected distances for the mock data (open circles). (Right) Recovery of slope as a function of true slope from weighted least-squares with projected distance as a proxy for true distance (open circles) and from our statistical framework (see text for details).}
\label{fig:spatial_distro}
\end{figure*}

In reality, a power-law distribution better describes the projected radial distribution of a typical GC system. In the left panel of Figure~\ref{fig:spatial_distro}, we compare the expected quantiles from a Rayleigh distribution versus the quantiles of the projected distances for our M87 sample (open circles). This figure demonstrates the extent the mock spatial distribution deviates from the assumption of our model. If the projected distances were drawn from a Rayleigh distribution, the theoretical quantiles versus the data quantiles would be a straight line. The Rayleigh distribution is not a significant deviation from the distribution of the observed projected distances in our sample.

Figure~\ref{fig:singlepop_viz} displays the joint probability distribution of all our parameters and data, $p(Z_{{\rm obs}, n}, r_{\perp, {\rm obs}, n}, Z_{{\rm true}, n}, \vec{r}_n, R, S, m, b, \sigma_n)$. Because the arrows indicate the conditional dependence among the parameters and data we use this to factorize the joint probability distribution of all our parameters into conditionally independent probability distributions to obtain, 

\begin{multline}
      p(Z_{{\rm obs}, n}, r_{\perp, {\rm obs}, n}, Z_{{\rm true}, n}, \vec{r}_n, R, S, m, b, \sigma_n) \propto \\ \prod_{n = 1}^N p(Z_{{\rm obs, n}} | Z_{{\rm true, n}}, S, \sigma_n) p(r_{\perp, obs, n} | \vec{r}_n) \times \\ \prod_{n = 1}^N  p(Z_{{\rm true}, n}  | m,  b, \vec{r}_n) p(\vec{r_n} | R)  p(\sigma_n)   \times  \\  p(S)p(R)p(m,b), 
\end{multline}

To test the efficacy of this model, we generated mock data from where the coordinates are drawn from a power-law that goes as $-2.5$, and each data point is randomly assigned $10\%$ to $55\%$ uncertainty. In the right panel of Figure~\ref{fig:spatial_distro} we compare how well we recover the true slope when using projected distance as a proxy for true distance and a weighted least-squares fit to get the gradients (open circles) to when how well we recover the slope when we marginalize over the scale length (closed circles). Over the range of slope values, the recoverability improves with the statistical de-projection. The difference in results between the two methods is starkest when the gradient is significant, while there is no difference in the recoverability when the gradient is consistent with being flat ($m=0$).

\begin{figure}
\includegraphics[width=0.4\textwidth]{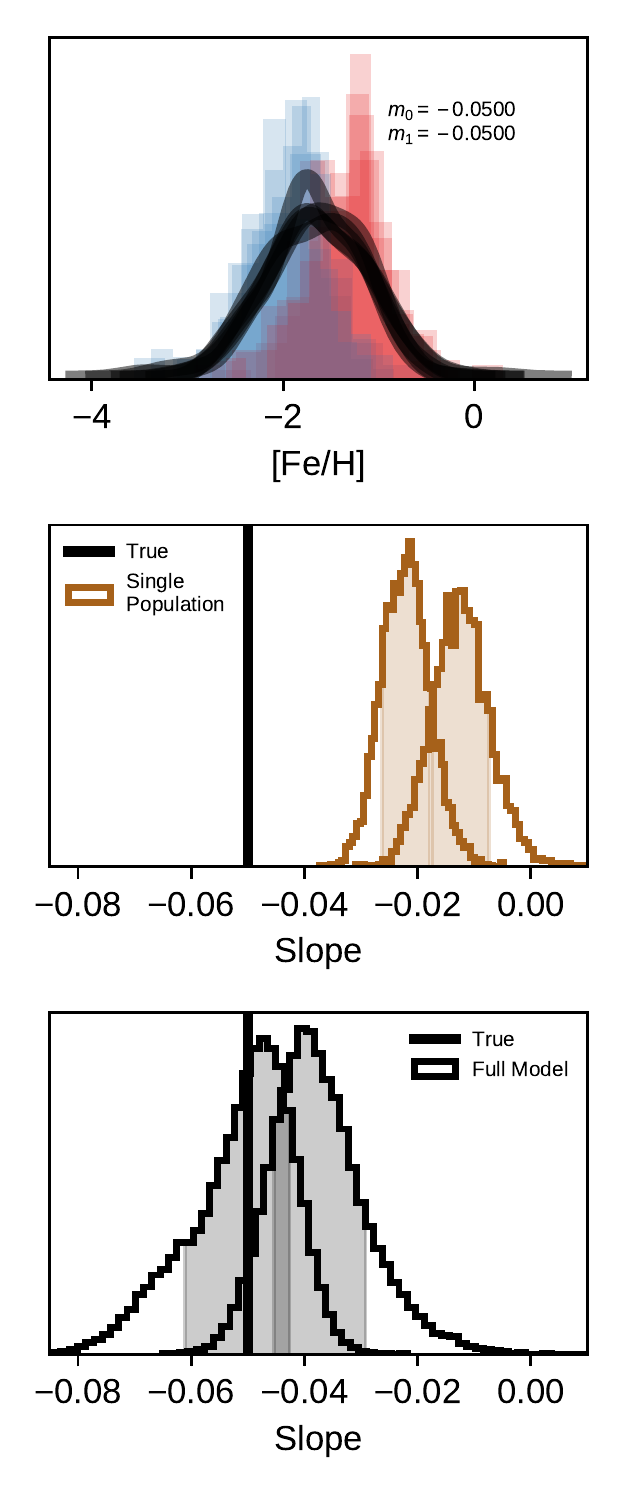}  
\caption{(Top) Metallicity distribution function (MDF) of 5 realizations of mock data generated from $m_0 = m_1 = -0.05$ and $b_0 = -0.4$ and $b_1 = -1.0$. Colored histograms show the true subpopulation separations and the black lines are the non-parametrically smoothed MDF of the combination of the subpopulations. (Middle) Demonstration of recovery of true slopes (black line) for when a the single population model from Section~\ref{sec:singlepop} is used on the subpopulations determined from a constant cut on [Fe/H] (brown)  Bands show the range between the 16th and 84th percentiles for all posteriors. (Bottom) Same as middle panel but now using the full hierarchical mixture model.
}
\label{fig:mock_t1}
\end{figure}

\subsection{Generalizing to multiple subpopulations}

\subsubsection{Effect of making cuts on the population}

\begin{figure}
\includegraphics[width=0.4\textwidth]{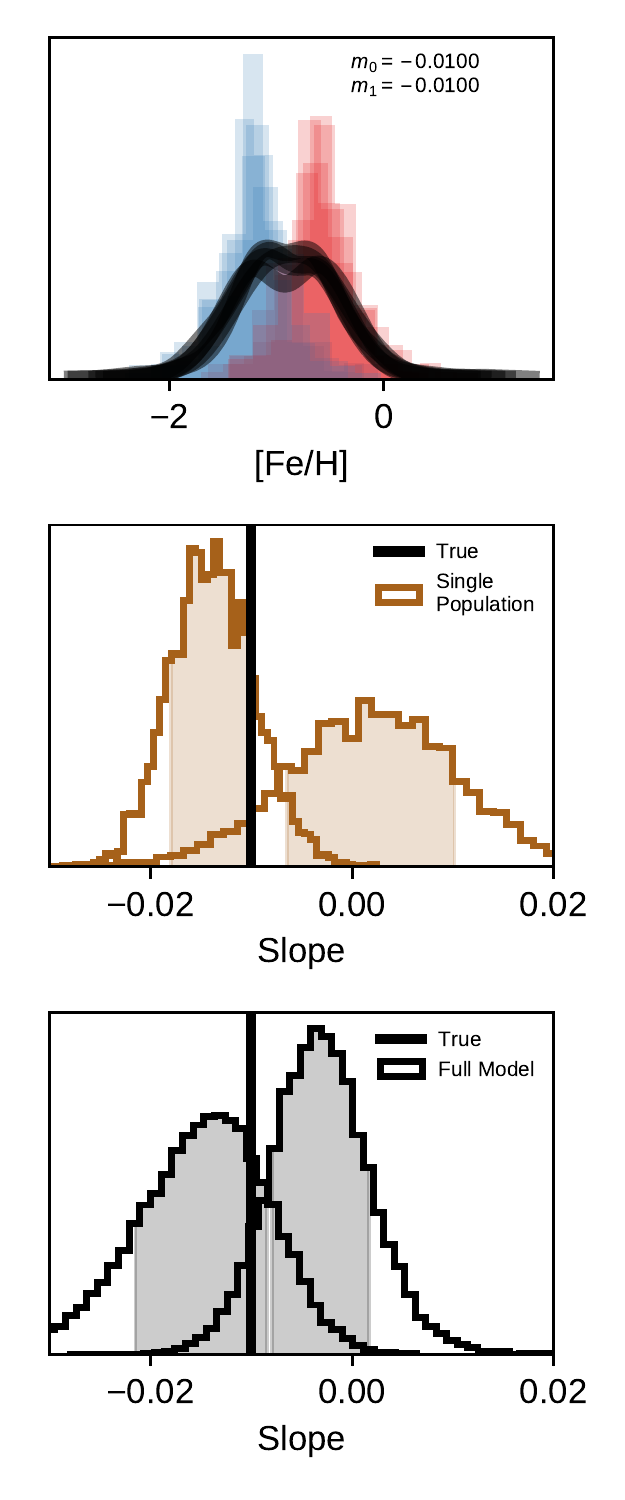}  
\caption{Same as Figure~\ref{fig:mock_t1} but for $m_0 = m_1 = -0.01$. 
}
\label{fig:mock_t3}
\end{figure}

In the previous section, we demonstrated the efficacy of a Bayesian linear regression approach relative to weighted least-squares to accurately recover the gradient parameters of a set of data points drawn from a particular line. A fundamental assumption in the method presented is the data points come from the {\it same population}. As previously described, however, GC systems are generally composed of subpopulations, and knowing how to separate the individual GCs of a system into the correct subpopulations is one of the most difficult steps towards characterizing GC systems.

In the top panels of Figures~\ref{fig:mock_t1} to \ref{fig:mock_t2} we show three versions of mock data, all generated from power law distributions and two underlying gradients. In all versions we use  $b_{{\rm true}, 0} = -0.4$ and  $b_{{\rm true}, 1} = -1.0$ and a variety of slope parameters:   $m_{{\rm true}, 0} = m_{{\rm true}, 1} = -0.05$ (Figure~\ref{fig:mock_t1}), $m_{{\rm true}, 0} = m_{{\rm true}, 1} = -0.01$ (Figure~\ref{fig:mock_t3}), and $m_{{\rm true}, 0} = -0.01$ and $m_{{\rm true}, 1} = -0.015$ (Figure~\ref{fig:mock_t2}). For each ``system'' we generated 5 realizations of mock data. 

For each set of mock data, we made constant cuts based on the MDFs to separate the populations, mimicking what one might do if they did not have {\it a priori} information on the different subpopulations. We fit each realization of the subsequent subpopulations with our Bayesian linear regression model presented in Section~\ref{sec:singlepop}, which we note is already an improvement over previously used methods, as demonstrated in Section 3.1.

In the middle panels of Figures~\ref{fig:mock_t1} to \ref{fig:mock_t2} we show how effective this method is by comparing the true slope values (black line) to the median of the posteriors for each realization of the mock data (brown histograms). Even with the improvements to the linear regression outlined in Section~\ref{sec:singlepop} the inferred slope values are not accurate, with the inferred slopes typically being flatter than the true slopes. We therefore need to generalize our single population model to account for the covariance between the gradient parameters and the subpopulation membership assignments to more accurately estimate both.

\subsubsection{A mixture model}

\citet{hogg2010a} discussed mixture models in the context of linear regression for the purposes of outlier rejection. Separating individual GCs into subpopulations is an equivalent problem. We model the system such that a given GC has $C$ number of subpopulations it could be assigned to through an identifier parameter $q_n$. Like \citet{hogg2010a}, we directly marginalize over the class membership of each GC by introducing a new parameter, the prior on $q_n$, $P_c \in [0, 1]$ such that $\sum_{c=1}^C P_c= 1$. The $P_c$ parameters are the mixture weights for each subpopulation and allow us to marginalize out the subpopulation identifiers.

In principle, we can fit for any number of subpopulations. In practice, however, throughout this work we specialize to the  $C=2$ case such that we model the mock and observed data as a bimodal distribution. We set a lower limit on $P_c$, $P_{\rm min} = 0.3$. Then, 

\begin{equation}
P_c = 
	\begin{cases}
	P_0 \sim {\rm Uniform}(P_{\rm min}, 1 - P_{\rm min}) \\
	P_1 \sim 1 - P_0	
	\end{cases}
\end{equation}

The structure otherwise remains the same as our single population model. We are able to transition our population parameters from the single population model to be a part of the mixture model because the parameters will exist for each mixture component (i.e., GC subpopulation). So we make a small adjustment to our notation:  $m_c$, $b_c$, ${\rm log}R_c$, and ${\rm log}S_c$ where the subscript refers to a given subpopulation. The joint probability distribution is then given by,

\begin{multline}
p(\alpha_C, \theta_n | x_n) \propto \\ \left(\prod^C_{c=1} p(R_c, m_c b_c, S_c) \right) \times \\ \prod^N_{n=1} \left(\sum^C_{c=1} P_c \times  p(Z_{{\rm obs}, n}, r_{\perp, {\rm obs}, n}, Z_{{\rm true}, n}, \vec{r}_n, R_c, S_c, m_c, b_c, \sigma_n) \right)
\end{multline}

The third term in this equation is what we factorized in Section 3.1 for the single population model. We show the graphical representation of the final hierarchical mixture model in Figure~\ref{fig:multipop_viz}.

\begin{figure}
\includegraphics[width=0.4\textwidth]{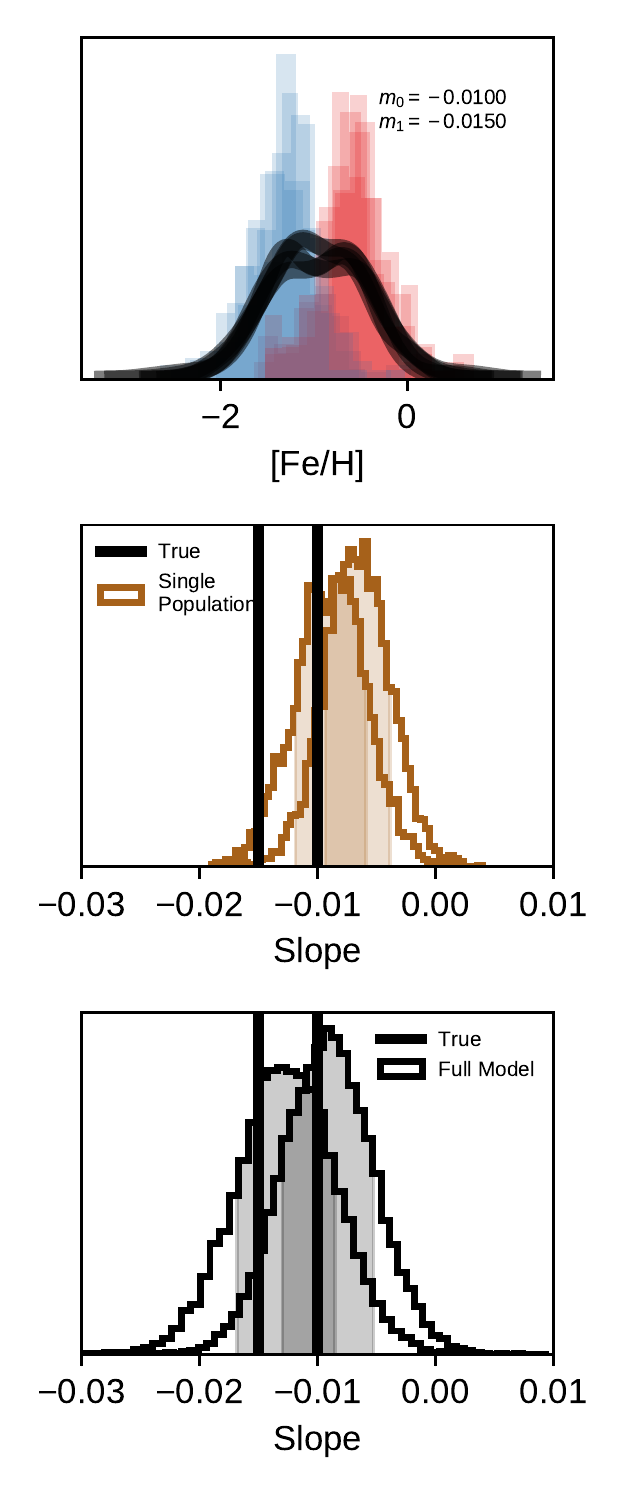}  
\caption{Same as Figure~\ref{fig:mock_t1} but for $m_0 = -0.01$ and  $m_1 = -0.015$. 
}
\label{fig:mock_t2}
\end{figure}

We specify our model with the probabilistic programming package \texttt{PyMC3} \citep[][]{pymc3}. \texttt{PyMC3} uses the Hamiltonian Monte Carlo (HMC) family of samplers. For this work in particular, we use the No-U-Turn Sampler \citep[NUTS,][]{NUTS}. HMC samplers are more efficient than the commonly used ensemble samplers because they do not rely on the current state to propose the next state \citep[for an introduction to HMC see][]{betancourt2017} and so they are the more appropriate choice for high-dimensional problems.  

The sampling efficiency of the HMC algorithm is highly sensitive to several tuning parameters. For this work, the most important tuning parameter is the mass matrix because our model parameters are highly covariant. If the mass matrix is not well-matched to the covariance of the posterior, both the step size will need to be decreased, and the number of steps increased, making it difficult to achieve convergence. 

\texttt{PyMC3} does not have a built-in way to optimize the mass matrix. We use the \texttt{exoplanet}\footnote{\url{https://exoplanet.dfm.io}} extensions to \texttt{PyMC3} to fit for a dense mass matrix during burn-in. We find values to to initialize the sampler via several steps: first, we fit a 1D mixture of Guassians on the metallicities while taking into account the uncertainties to get an initial guess of the class membership for each observation, and then, second, we fit a linear model to the project metallicity gradients for each subpopulation to find initial guesses for the intercepts and slopes. With this approach, we obtain a converged model based on the Gelman-Rubin statistic for each parameter, $\hat{R}$, (where $\hat{R} > 1$ indicates the chains have not converged).

In the bottom panels of Figures~\ref{fig:mock_t1} to \ref{fig:mock_t2} we show the inferred slope posteriors (black histograms) to demonstrate the efficacy of this method. In all cases the recovery is better when using the full model, with the biggest improvement made in the case where the two subpopulations are most well-mixed in the MDF (Figure~\ref{fig:mock_t1}). 

The full model accurately recovers the different slopes in Figure~\ref{fig:mock_t2} within $1\sigma$ uncertainty but cannot distinguish the gradients as different at a statistically significant level. This is still an improvement over existing methods, but, in the context of GC subpopulations, the ability to discern any gradient differences is essential for understanding how potentially similar the assembly histories of the different subpopulations (see Section 5 for more discussion on this). Improving the precision of the subpopulation parameters will be the subject of future work. 

\begin{figure}
\includegraphics[width=0.5\textwidth]{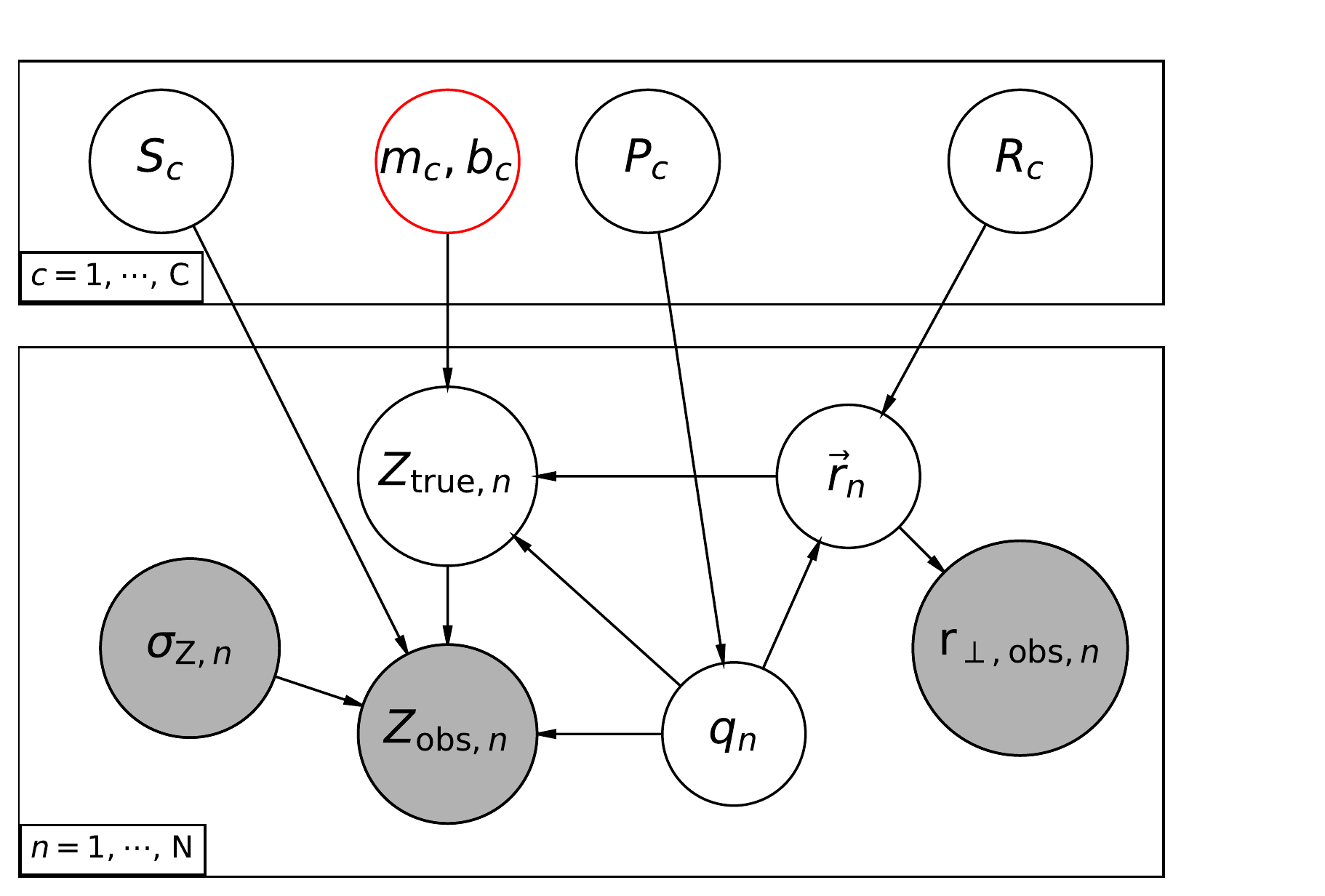}
\caption{Similar to Figure~\ref{fig:singlepop_viz} but now for our final hierarchical mixture model (see text for details). Now there is a second plate around are population parameters which indicates these parameters are determined} for all subpopulations in our sample ($c=1,...,C$) and we have subpopulation identifiers, $q_n$ set by the prior $P_c$.
\label{fig:multipop_viz}
\end{figure}

\section{Results}
\label{sec:results}

\subsection{Radial metallicity gradients}

In this work, we have two spectroscopic data sets for the GC system: the Keck/LRIS sample and the MMT/Hectospec sample. The former covers a radial range of $\sim 7 - 27$ kpc while the latter spans $\sim 14 - 142$ kpc.  Previous kinematical analyses of GCs and planetary nebula show signs of a transition at $\sim 40$ to $50$ kpc, which may be related to a recent accretion event \citep[][]{romanowsky2012, longobardi2015, zhang2015}. Photometric surveys have also shown in M87, and other massive ETGs, that blue GCs begin to dominate at large radii. To measure the metallicity gradients, we split our sample at $40$ kpc. The inner halo sample consists mostly of the Keck/LRIS data with a small fraction coming from the MMT/Hectospec data. The outer halo sample consists completely of MMT/Hectospec data. 

\begin{figure}
\includegraphics[width=0.5\textwidth]{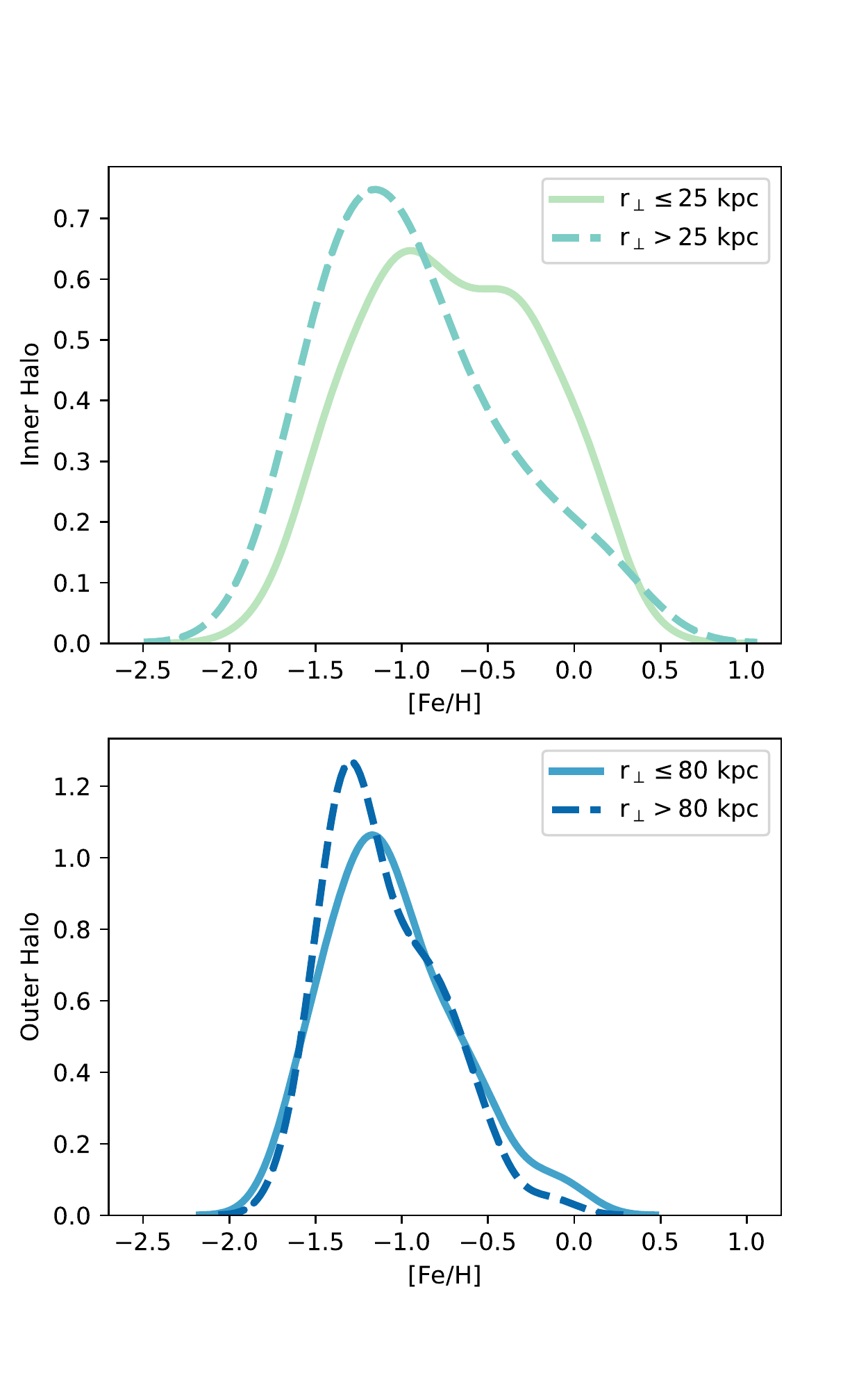}  
\caption{Empirical demonstration of gradients for inner (top) and outer (bottom) halos. In each panel we show the metallicity distribution function of the data set broken into two radial bins. Both the inner and outer halos show evidence of multiple subpopulations from their MDFs and a slight gradient.
}
\label{fig:empirical_demo}
\end{figure}

Before discussing the results from fitting the model to the individual [Fe/H] measurements, we first establish our expecations empirically in Figure~\ref{fig:empirical_demo}. In the top panel we show two metallicity distribution functions (MDF) for the Keck/LRIS sample, one for the inner part of the dataset (light green line, $r_\perp \leq 25 $ kpc) and one for the outer part (dark green dashed-line, $r_\perp > 25$ kpc). There are two distinct peaks in the inner MDF, while the outer MDF has a less significant second, metal-rich peak. In the outer bin, the metal-poor peak shifts noticeably from the inner metal-poor peak.

In the bottom panel we do the same demonstration for the outer halo. Bimodality is not as clearly seen in the outer halo sample as it is in the inner halo but there is a distinct negative shift from the main peak from the inner bin to the outer bin. The lack of clear bimodality could be a result of the MMT/Hectospec sample having far fewer red GCs than blue GCs and is consistent with the findings for other BCGs \citep[see, for example,][]{harris2017a}.

\begin{figure}
\includegraphics[width=0.5\textwidth]{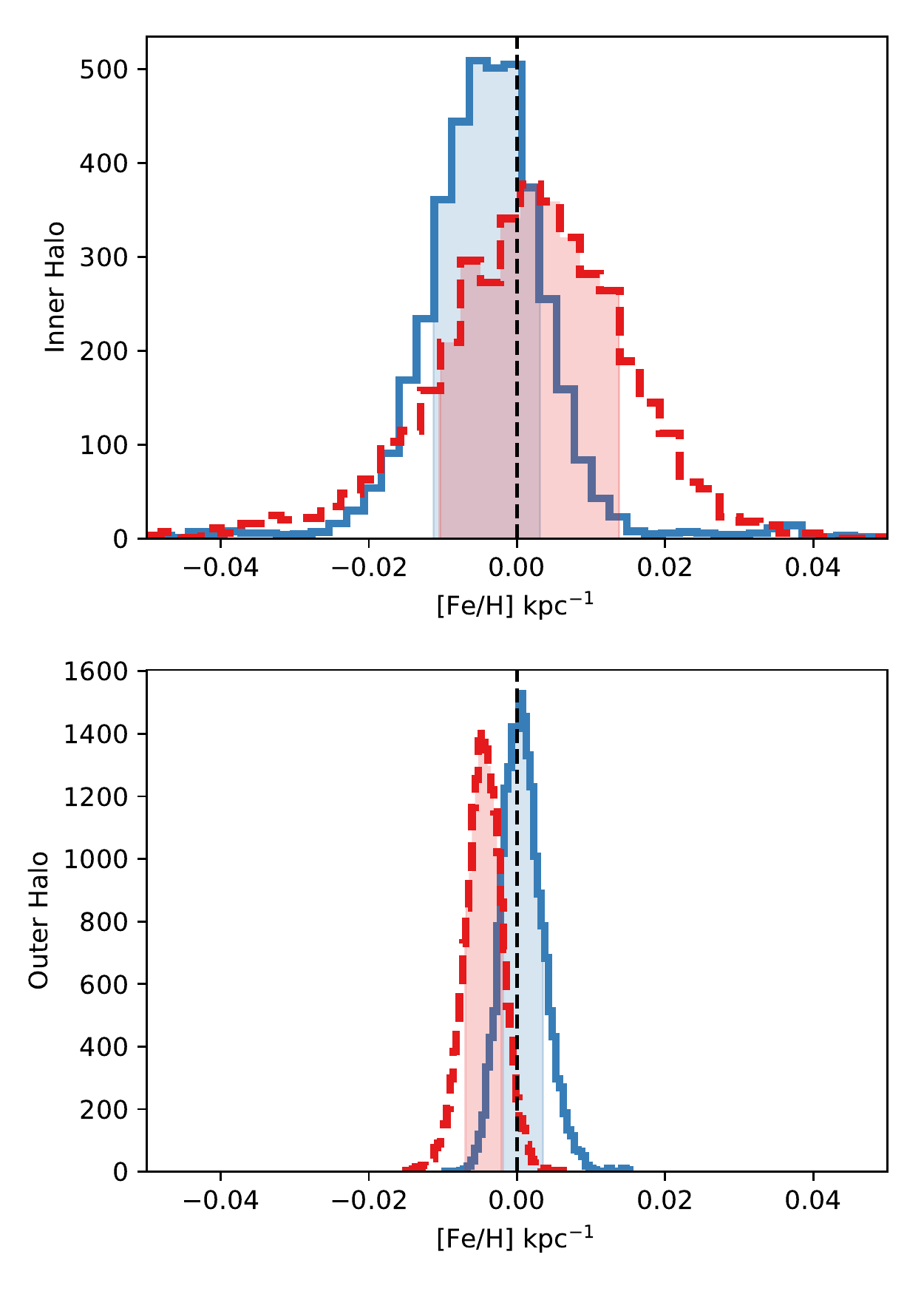}
\caption{Comparing the posteriors on the slopes for the metal-poor (blue) and metal-rich (red) subpopulations for the inner (top) and outer data (bottom) halos. The 1$\sigma$ uncertainty in each posterior is shown in the colored bands and a flat gradient is marked (black dashed line). 
}
\label{fig:slope_posteriors}
\end{figure}

For the modeling, we initialized the MCMC chains in the same manner as the mock data and modeled the data as composed of two subpopulations for both the inner and outer halos. In Figure~\ref{fig:slope_posteriors} we show the posteriors of slope values for the metal-poor (blue) and metal-rich populations for the inner halo (top panel) and the outer halo (bottom panel). The 16th and 84th percentiles are marked by the colored bands. In each panel a flat gradient is marked with the black dashed line.

The uncertainty on the slope measurements is significantly larger for the inner halo than the outer halo measurements even though the [Fe/H] uncertainty is $\sim 20\%$ higher for the MMT/Hectospec data. The large uncertainty in the inner halo could be a result of the comparatively non-uniform coverage in $r_\perp$ for the inner halo sample, we get less information from each individual measurement in the inner halo than the outer halo. For the inner halo, both the metal-rich and metal-poor slopes are consistent with a flat gradient and are statistically consistent with one another. In the outer halo, the metal-poor slope is consistent with a flat gradient while the metal-rich slope is slightly negative.

\begin{figure*}
\includegraphics[width=1\textwidth]{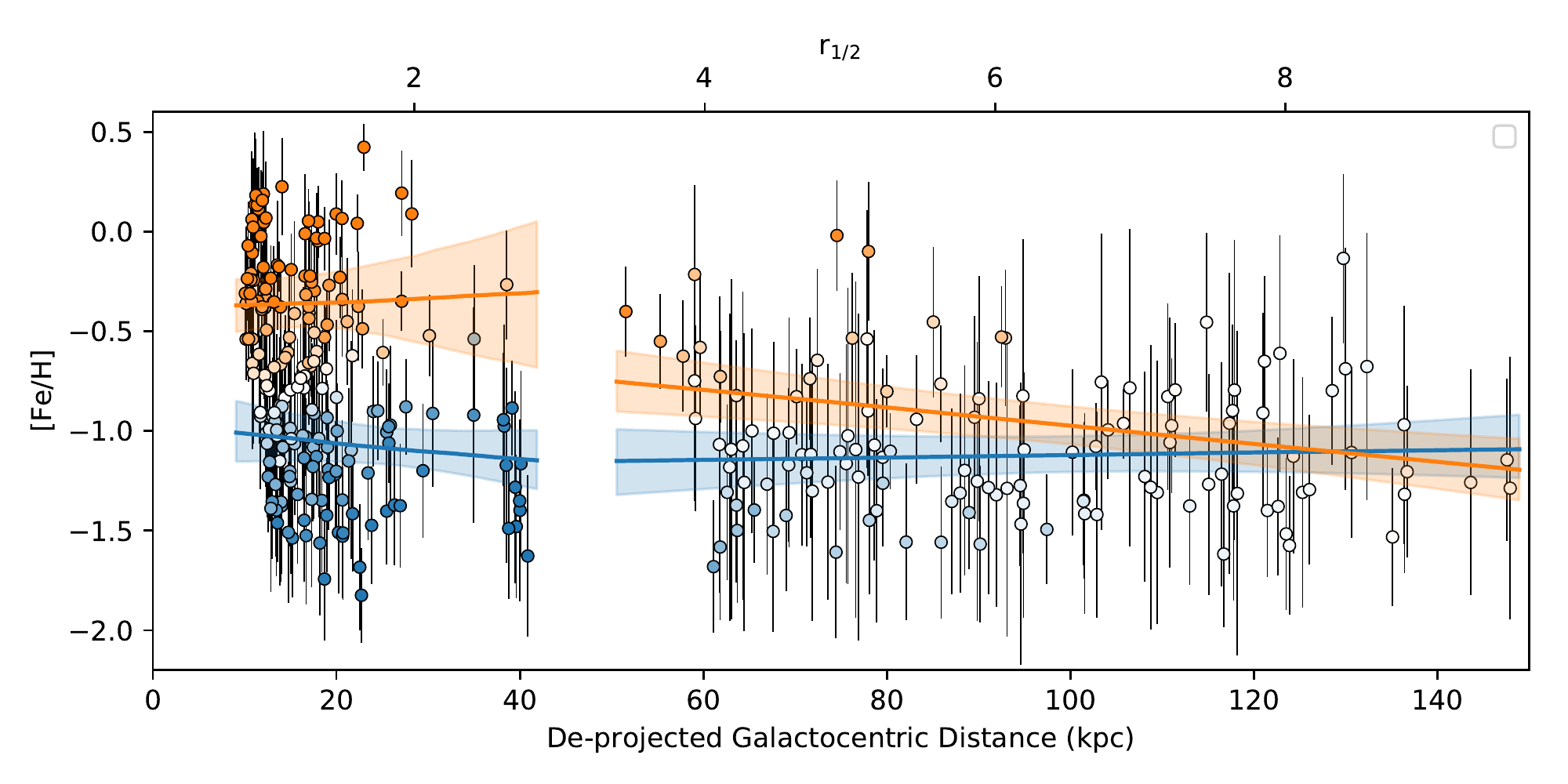}
\caption{Radial metallicity gradients of the subpopulations with respect to the de-projected distances. The circles show the [Fe/H] measurements. They are colored by subpopulation assignment and the opacity of the individual points is scaled by certainty of that subpopulation assignment, with white indicating the assignment is highly uncertain. We show posterior median (solid lines) and the range encompassed by the 16th and 84th percentiles (bands) of the gradient distributions.
}
\label{fig:full_gradient}
\end{figure*}

In Figure~\ref{fig:full_gradient} we show the metallicities as a function of the de-projected distances where the data points are colored by subpopulation assignment. The opacity of the individual points is scaled by certainty of that subpopulation assignment, with white indicating the assignment is highly uncertain. We show posterior median (solid lines) and the range encompassed by the 16th and 84th percentiles (bands) of the gradient distributions. Even though the gradient parameters are more uncertain in the inner halo, the subpopulation membership assignments are more certain than in the outer halo population because there are fewer metal-rich GCs and the metallicity separation between the subpopulations is smaller. 

\begin{deluxetable*}{lllllccc}
\tabletypesize{\footnotesize}
\tablecolumns{8}
\tablewidth{0.9\columnwidth}
\tablecaption{Summary of gradient parameters and MDF characteristics for the subpopulations of the M87 GC system. \label{table:results_summary}}
\tablehead{
& \colhead{$m$}  & \colhead{$\sigma_m$} & \colhead{$b$} & \colhead{$\sigma_b$} &  & \colhead{MDF} &  \\
 &  &  &  & & \colhead{16th} & \colhead{50th} & \colhead{84th}
}
\startdata
\centering
Inner Halo       &     &          &     &          &              \\
Metal-Poor &  $-0.004$   &  $0.010$   & $-0.957$ & $0.224$     & $-1.43$ & $-1.15$  & $-0.91$         \\
Metal-Rich &  $-0.001$   &  $0.014$   & $-0.384$ & $0.216$      & $-0.64$ & $-0.34$  & $+0.05$          \\
\hline
Outer Halo   &     &          &     &          &              \\
Metal-Poor &  $+0.001$  &  $0.003$   & $-1.196$ & $0.287$     & $-1.42$ & $-1.24$  & $-0.96$        \\
Metal-Rich &  $-0.005$   &  $0.003$   & $-0.522$  & $0.268$     & $-0.83$ & $-0.64$  & $-0.45$       
\enddata
\end{deluxetable*}

\subsection{Characteristics of the subpopulations}

\begin{figure*}
\includegraphics[width=1.0\textwidth]{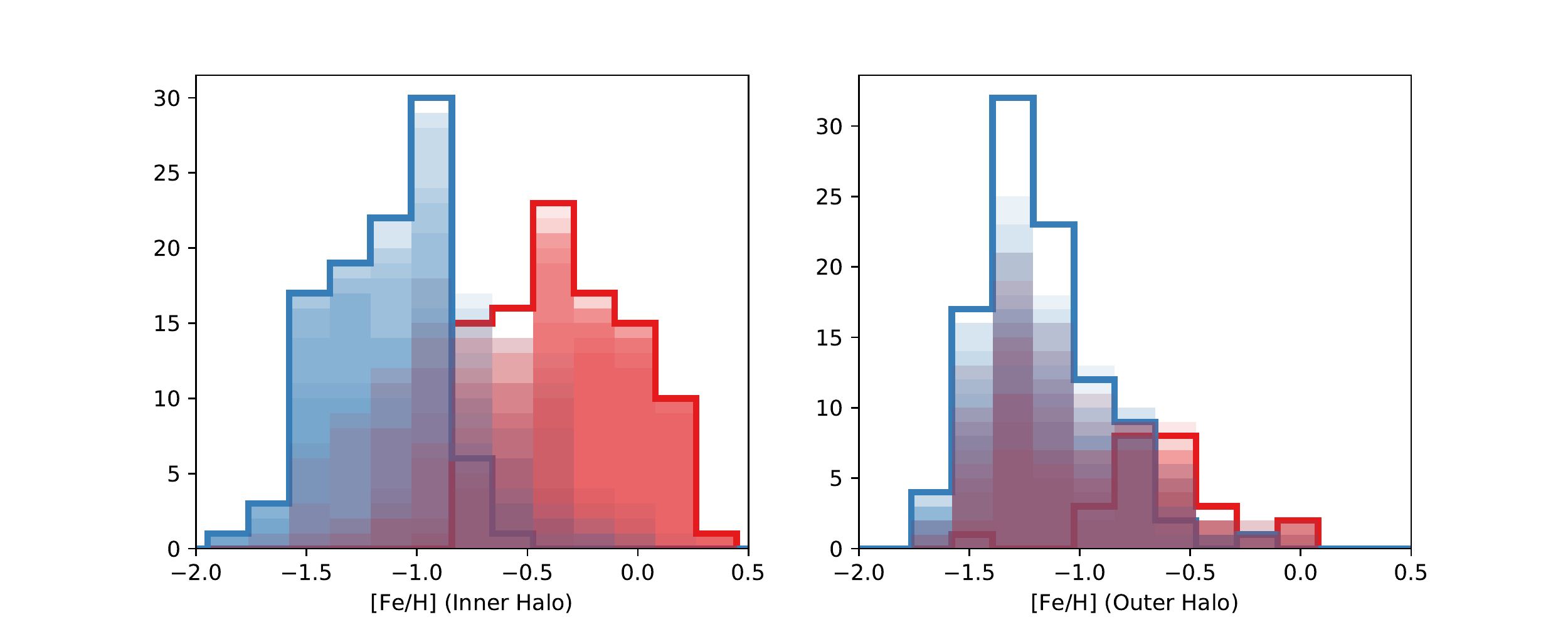}
\caption{(Left) MDF for the metal-rich (red) and the metal-poor GCs (blue) in the inner halo.The solid blue and red lines show the posterior median of the subpopulation assignments of the individual GCs. The filled-in blue and red histograms represent how the uncertainty in the subpopulation assignments (see text for details) propagates to uncertainty in the MDF. (Right) Same as left but for the outer halo GCs.
}
\label{fig:mdf}
\end{figure*}

In Figure~\ref{fig:mdf} we compare the metallicity distribution functions (MDFs) for the metal-poor GCs (blue) and metal-rich GCs (red) for the inner halo GCs (left) and outer halo GCs (right). The solid lines show the result of using the posterior median of the subpopulation assignments of the individual GCs. We represent how the uncertainty in the subpopulation assignment affects the MDF by plotting the result of selecting class labels using a random number generated by the probability of the cluster-subpopulation pair for 10 random samples from the posterior (solid histograms). 

To check the results of our model we compare Figure~\ref{fig:empirical_demo} with Figure~\ref{fig:mdf}. Figure~\ref{fig:empirical_demo} indicates that the two subpopulations should be of about equal size for the inner halo sample and that the metal-poor GCs would be a larger population in the outer halo sample. Even though the subpopulation membership assignments are much less certain for the outer halo sample, we see that the model assigns significantly fewer metal-rich GCs in the outer halo. This picture is overall consistent with our broad understanding that with increasing galactocentric distance there will be more metal-poor GCs relative to metal-rich. 
 
In Table~\ref{table:results_summary} we summarize the gradient parameters and the characteristics of the MDFs for the subpopulations.

\begin{figure}
\includegraphics[width=0.5\textwidth]{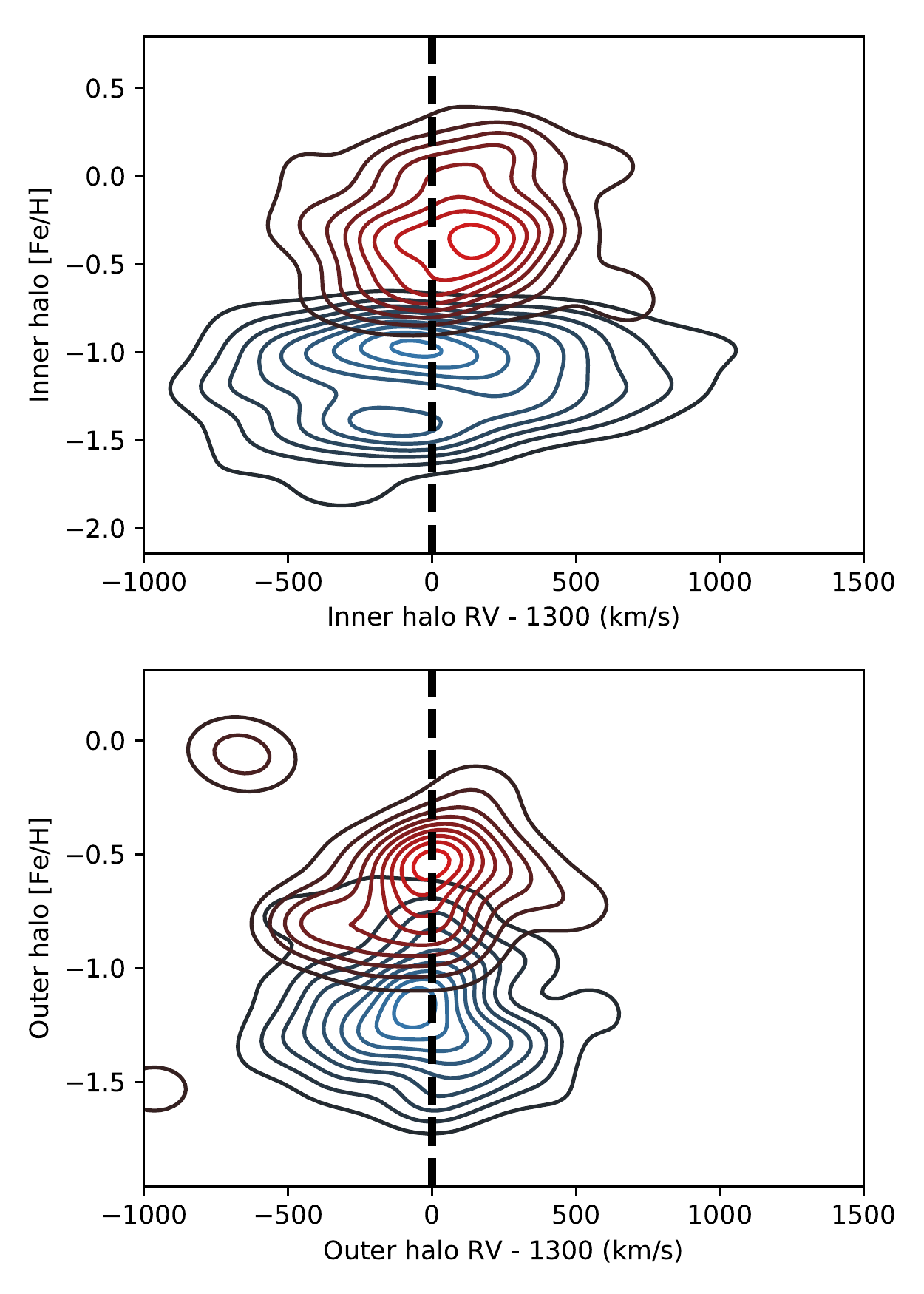}
\caption{[Fe/H] vs. radial velocity for the inner halo sample (top) and the outer halo sample (bottom).
}
\label{fig:rv}
\end{figure}

In the Milky Way GC system, the metallicity subpopulations are associated with different spatial and kinematical components of the Galaxy itself \citep{zinn1985}. In Figure \ref{fig:rv} we show how well this pattern holds for M87 by examining the chemodynamics of the subpopulations in [Fe/H]--radial velocity space for the inner halo (top) and outer halo (bottom).
For the inner halo, our modeled metallicity subpopulations correspond to differences in the radial velocity distributions. The metal-rich subpopulation has significantly less radial velocity dispersion than the metal-poor subpopulation. 

Unlike the inner halo, there does not seem to be a correspondence between metallicity subpopulation and differences in the kinematic properties of the subpopulations for the outer halo. The subpopulations in the outer halo have a radial velocity dispersion similar to the inner halo metal-rich subpopulation.

\subsection{Abundance patterns}

\begin{figure*}
\includegraphics[width=1\textwidth]{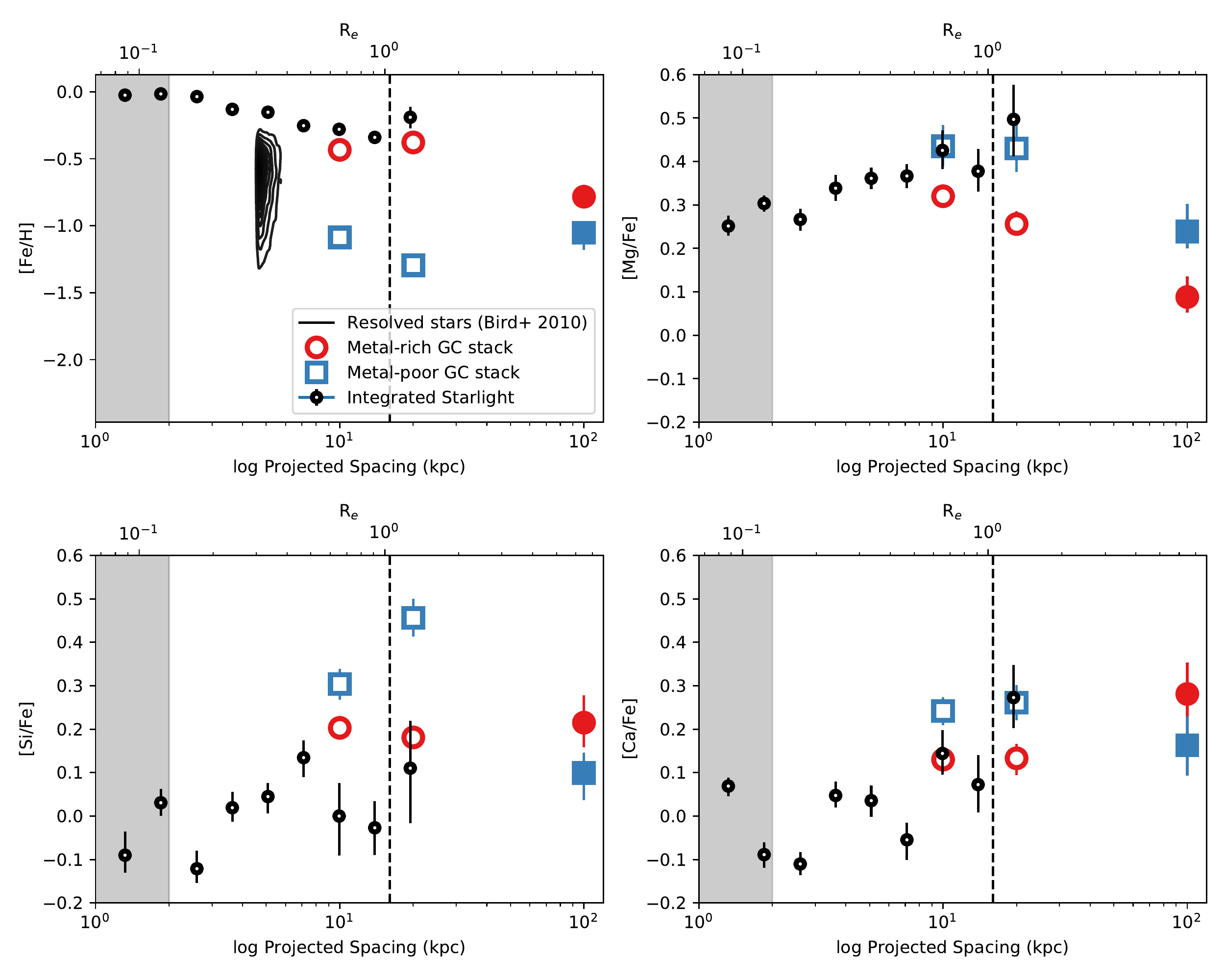}
\caption{Stellar population radial profiles for [Fe/H] and a variety of $\alpha$-elements as derived from full spectrum fitting to the \citet{murphy2011} spectroscopy of the M87 galaxy light (black), the metal-rich GC stacks (red circles), and the metal-poor GC stacks (blue squares). The grey band indicates $R_{\rm gal} \leq 2.0$ kpc, i.e., the central-most region where massive ETGs display many exotic stellar population characteristics, and the dashed line indicates $\sim1R_e$. 
}
\label{fig:alpha_contact_sheet}
\end{figure*}

In Figure~\ref{fig:alpha_contact_sheet} we show the stellar population radial profiles for M87 from our fits to the \citet{murphy2011} sample (black circles) for [Fe/H] (upper-left) and a variety of $\alpha$-elements. The M87 starlight shows a declining [Fe/H] profile and slightly positive profiles for [Mg/Fe] (upper-right), [Si/Fe] (lower-left), and [Ca/Fe] (lower-right). The [Fe/H] gradients are consistent with previous work that have studied stellar population gradients in massive ETGs \citep[e.g.][]{greene2015, vd2017a, gu2018a}. These previous studies typically found a flatter [Mg/Fe] than what we present here but it is not a substantial difference. 

In the upper-left panel we also show [Fe/H] estimates from deep broadband photometry of resolved stars in M87 \citep[contours,][]{bird2010}. This comparison shows a metal-poor field star population which is not probed by the stellar population models. The comparison of the resolved stars to the integrated light demonstrates the limitations inherent in studies using integrated light and the need for the inclusion of the GC population in the analysis. 

To obtain abundance information for the GCs we have to stack the individual spectra since the majority of the Keck/LRIS and MMT/Hectospec spectra have too low-S/N to reliably extract abundance information. Stacking the GC spectra is made difficult by the need to separate the sample by subpopulation as it is expected the different subpopulations will have different origins and, thus, different abundance patterns. We demonstrated in Section 3 the importance of using a HBM framework for making accurate determinations of subpopulation membership of the individual GCs, which help make more physically-appropriate stacks. Additionally, we can take advantage of having made probabilistic determinations of subpopulation membership for the individual GCs. In Figure~\ref{fig:mdf} we demonstrated how uncertainty in the subpopulation memberships propagated to the MDF of the GC system. In the same manner, we can propagate that uncertainty to our stacks and abundance information. 

In the same manner we used for the Mitchell data, we made four inner halo stacks, binning by metal-rich and metal-poor and then further separating the GCs at a radius at $16$ kpc, and two stacks for the outer halo only binning by metal-rich and metal-poor. We made ten iterations for each subpopulation, determined by different draws from the posterior for different subpopulation assignments for each GC (same draws that are shown in Figure~\ref{fig:mdf}). The stacked GC spectra have a typical S/N of $\sim 100 - 150/{\rm \AA}$.

Each version of each subpopulation stack was fitted using \texttt{alf} in the same manner as the Mitchell data. For each parameter of interest, we computed the 16th, 50th, and 84th percentiles of the posteriors for each fit. In Figure~\ref{fig:alpha_contact_sheet} we show the results for the metal-rich stacks (red circles) and metal-poor stacks (blue squares). The inner halo stacks are open symbols and the outer halo stacks are filled.

While we note that Figure~\ref{fig:alpha_contact_sheet} cannot be directly compared to Figure~\ref{fig:full_gradient} because we have moved from de-projected to projected distances, broadly the [Fe/H] gradient measured from the metal-rich inner halo stacks is consistent with the flat gradient measured from the individual [Fe/H] measurements. For the inner halo metal-poor stacks we find a slightly negative gradient from the stack measurements that differs from the flat gradient shown in Figure~\ref{fig:full_gradient}. The likely cause of this difference is the non-uniform sampling of the inner halo GCs in galactocentric radius. For the individual [Fe/H] measurements the MMT/Hectospec sample provides the only coverage past $\sim 27$ kpc and the more metal rich measurements ($\sim -1.0$) seem to be enough to keep the gradient nearly flat. However, for the stacks there are fewer of these comparatively metal-rich GCs than metal-poor so their contribution to the stack is not as important.

The inner halo metal-rich stacks have a similar metallicity to the M87 starlight in the same region while the metal-poor stacks are less metal rich by $\sim 1$ dex. For [Mg/Fe] the inner halo metal-rich stacks are less enhanced than the galaxy light while the metal-poor stacks have similar abundances for the galaxy light. For both outer halo stacks, there is a precipitous drop in Mg-enhancement. For [Si/Fe], the inner halo metal-poor stacks are enhanced relative to the galaxy light and metal-rich stacks, which have similar values. There is a noticeable drop in enhancement for the outer halo metal-poor stack while the outer halo metal-rich stack is consistent with the inner halo measurements. The [Ca/Fe] measurements are consistent between the galaxy light and all the inner halo stacks, and the outer halo stacks are consistent with the inner halo measurements.

\begin{figure*}
\includegraphics[width=1\textwidth]{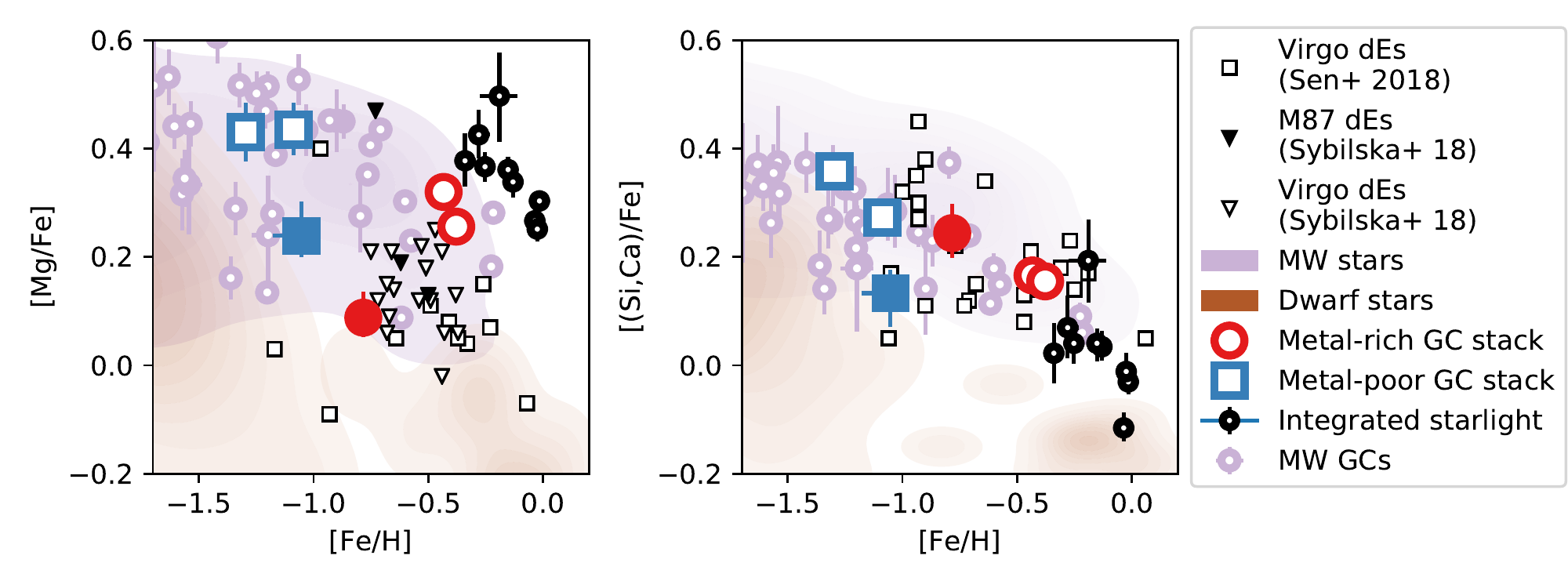}
\caption{(Left) [Mg/Fe] vs. [Fe/H] for the GC stacks and M87 (symbols same as previous figure) For M87 the two measurements we have that are within $2$ kpc are filled in. Also plotted is the kernel density estimate of the Milky Way field stars (purple) and the field stars from the Milky Way dwarf satellite population (brown), Virgo dEs (open triangles and squares), M87 dEs (objects within 300 kpc of M87, closed triangles), and integrated light measurements of Milky Way GCs (purple circles). (Right) Same as left panel but for the median of [Si/Fe] and [Ca/Fe]. 
}
\label{fig:alpha_abundances}
\end{figure*}

In the left panel of Figure~\ref{fig:alpha_abundances} we show [Mg/Fe] vs. [Fe/H] for the GC stacks and the galaxy data (colors and symbols same as previous figure). We also show the abundances for the Milky Way stars \citep[purple cloud; from the JINAbase][see detailed references in Appendix \ref{ap:jinarefs}]{jinabase}, stars in dwarf galaxies around the Milky Way \citep[brown cloud; JINAbase and ][]{bonifacio2004}, and Milky Way GCs fitted from \citet{schiavon2005} \citep[purple circles; see][for details on these fits]{villaume2019}. Also displayed are the abundances for dwarf ellipticals (dEs) in Virgo from two different studies, \citet{sen2018} (open squares) and \citet{sybilska2018} (triangles). For the \citet{sybilska2018} sample we differentiate between ``M87 dEs'' (closed) and ``Virgo dEs'' (open) with a cut at $300$ kpc from M87  \citep[distances from][]{peng2008}. We were motivated by the results from \citet{liu2016} which showed a transition in [Mg/Fe] in the dwarf elliptical population at this distance.

In the right panel of Figure~\ref{fig:alpha_abundances} we show the median values for each object of [Si/Fe] and [Ca/Fe], except for \citet{sen2018}, who only measured [Ca/Fe]. \citet{sybilska2018} is not shown on this panel as they only measured [Mg/Fe]. All symbols are the same as the left panel. The outer halo metal-rich stack is enhanced in [(Si,Ca)/Fe] but much closer to solar in [Mg/Fe].

\begin{figure*}
\includegraphics[width=1\textwidth]{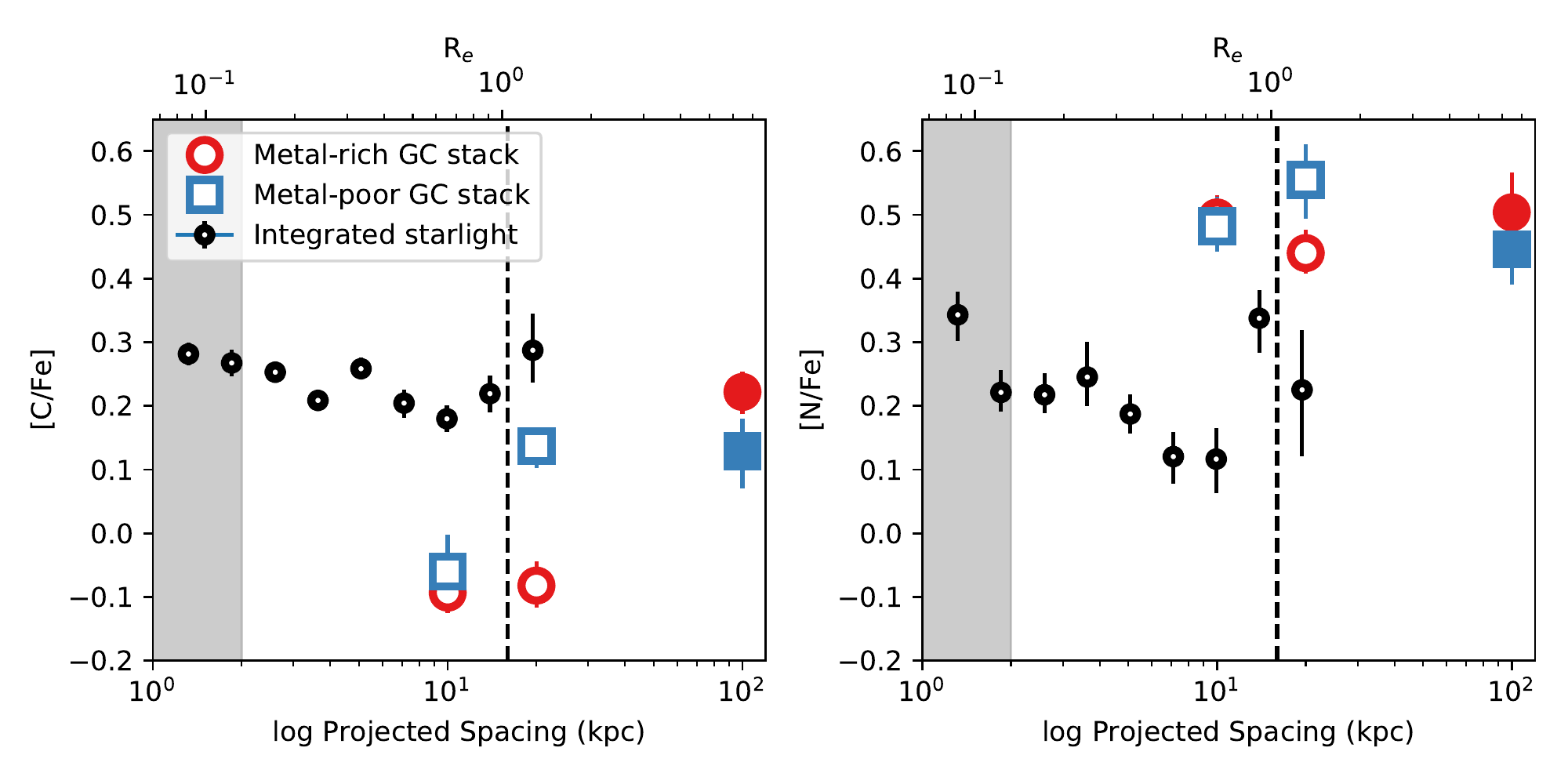}
\caption{Same as Figure~\ref{fig:alpha_contact_sheet} but for  [C/Fe] (left) and [N/Fe] (middle) (right).
}
\label{fig:light_contact_sheet}
\end{figure*}

In Figure~\ref{fig:light_contact_sheet} we show a similar figure to Figure~\ref{fig:alpha_contact_sheet} but now for light elements: radial profiles for [C/Fe] (left) and [N/Fe] (right). For the M87 starlight we see enhanced [C/Fe] and [N/Fe] values and with negative radial gradients for both abundances. The GCs as a whole are less enhanced than the galaxy starlight for [C/Fe] but more enhanced in [N/Fe]. 

\section{Discussion}
\label{sec:discussion}

\subsection{The formation of the inner halo} 
\label{sec:inner_halo}

Piecing together how the inner halo ($< 40$ kpc) formed is complicated by the fact that it is a mix of {\it in-} and {\it ex-situ} stellar populations of unknown proportions. 
Decomposing the whole population into these components from observations cannot be quantitatively done with integrated galaxy starlight alone.
With the GC system we have discrete tracers of near-simple stellar populations that overlap with and extend our coverage of the galaxy field star population, providing additional insight into how this region formed.

The red GC populations in massive ETGs have long been thought to have formed along with the original galaxy because they follow the field star density profiles and kinematics \citep[see][for M87 in particular]{strader2011b}.
Until this work, however, a direct metallicity comparison at the same galactocentric radius has not been done. 
We have established that the metal-rich GCs have an average [Fe/H]$\sim -0.4$ and [Mg/Fe]$\sim +0.15$ (Figures~\ref{fig:alpha_contact_sheet} and \ref{fig:alpha_abundances}), similar to the galaxy field star population ([Fe/H]~$\sim -0.3$ and [Mg/Fe]~$\sim +0.40$) over the same radial extent, indicating a common origin of the two populations.

In Figure~\ref{fig:alpha_contact_sheet} we show the radial gradients for [Fe/H] (upper-left panel) and various $\alpha$ elements measured from the integrated light of M87 (black circles). We find that the [Fe/H] gradient for the field star population is flat within the inner 2 kpc and then steepens to a negative gradient. This negative gradient is paralleled by {\it rising} gradients in all of the $\alpha$ elements.
This trend is characteristic of the populations seen in other massive galaxies \citep[][]{gu2018b, greene2013} and can be viewed as consistent with the second phase of the ``two-phase'' formation framework \citep{oser2010} within a preferentially quenched environment \citep{liu2016}. 

It would follow from this scenario that the bulk of the metal-rich GCs came in from mergers.
However, the measurements from the GC stacks indicate that the [Mg/Fe] values for the metal-rich GCs decline with radius. The population that then presumably brought in the metal-rich GCs would be diluting the [Mg/Fe] and depressing the gradient, rather than contributing to its rise. On the other hand, the metal-poor GCs are very Mg-enhanced.

A negative gradient is expected from the kind of minor mergers that would bring metal-poor GCs into M87. Major mergers, which would bring in metal-rich GCs, can flatten gradients \citep[e.g.][]{taylor2017}.
Our current measurements show that the metal-poor subpopulation gradient is skewed negative but is consistent with the metal-rich (and with a flat gradient; Figure  \ref{fig:slope_posteriors}) within the $1\sigma$ uncertainties. It would presumably follow then that the metal-poor GC population was affected by the same processes that flattened the metal-rich GC gradient, i.e., that the metal-poor population was already in place by the time the major mergers began and, thus, the low-mass satellites were preferentially accreted early. To the best of our knowledge, this is not something that has been examined in cosmological simulations of galaxy evolution. However, we can use the additional constraint of the metallicity gradient of the galaxy light. If all the low-mass satellite galaxies came in early, we would expect that the stellar metallicity gradient should be similarly flat to the GC subpopulation gradients and so we conclude that scenario is not likely. 

An alternative explanation is that the metal-poor population is not entirely {\it ex-situ}. \citet{mandelker2018} described a scenario in which metal-poor GCs form {\it in-situ} in massive halos (i.e., viral masses a few times $10^{10} - 10^{11} M_\odot$) at high-redshift ($z\sim6$) as a result of fragmentation of the unstable cold gas filaments accreting on the forming galaxy. To the best of our knowledge, at this redshift the circumgalactic gas does not evolve \citep[see Figure 6 in][]{robert2019} and so there would be no mass--metallicity relation, and thus, the population of GCs that stem from this scenario would have a flat metallicity gradient. An {\it in-situ} metal-poor population commingling with the {\it ex-situ} metal-poor GC population could then ``dilute'' the overall ``metal-poor'' subpopulation gradient we measure.

A related issue regarding the origins of the inner halo metal-poor GC population is that cosmological simulations of galaxy evolution do not predict such a low-metallicity population. Cosmological simulations predict that the Milky Way-mass (mass ratio$\sim$1:5) galaxies are the primary building blocks of the stellar halos of massive galaxies \citep[][]{oser2012, pillepich2018}, while the typical mass implied by the median metallicity of the metal-poor GC population is much lower, $M_* \sim 10^8$ \citep{kirby2013}.
This makes the metal-poor GCs too metal-poor to fit this framework, even though in \citet{villaume2019} we established that the metal-poor GCs in the inner halo of M87 are $\sim 0.4$ dex more metal rich than previously thought. The mass ratios of the mergers suggested by this metal-poor population are not inconsistent with analytic models of merger-driven BCG expansion from high-redshift to present day \citep[see equation 4 in][]{naab2009}.

\subsubsection{Metal production in the field and GC populations}

Including the GC system in the analysis helps establish important benchmarks -- the existence of the $\alpha$-enhanced, metal-poor halo and the flat (albeit not necessarily the same) metallicity gradients of the GC subpopulations  -- that help us make qualitative advancements in our understanding of the assembly history of M87. However, until we can make more precise determinations of the {\it in-} and {\it ex-situ} populations we necessarily have to be agnostic towards the specifics of the {\it in-situ} star-formation in M87. This, however, is a particularly important process to understand because of the unexplained, exotic properties of the stellar populations in the innermost regions of the most massive ETGs. For example, the unexpected excess in ultraviolet (UV) flux within \citep[e.g.,][]{code1979} and the bottom-heavy initial mass functions up to $1R_e$ \citep[][]{vd2017b}. For M87 in particular, \citet{sarzi2018} measured a bottom-heavy IMF out to $\sim4$ kpc.
Constraining the nature of this initial phase of ETG formation will likely clarify the star formation processes that can give rise to these characteristics. 

The radial gradients of the various abundances shown in Figures \ref{fig:alpha_contact_sheet} and \ref{fig:light_contact_sheet} contain a wealth of information which can help falsify various formation scenarios. In this paper, we focus specifically on the scenario which postulates that there is a significant fraction of stars from dissolved metal-rich GCs in the center of M87 which explains the UV upturn \citep{goudfrooij2018}. One explicit prediction from this scenario is a negative radial gradient in [N/Fe] in the galaxy starlight, which we show to be case in the right panel of Figure \ref{fig:light_contact_sheet}. Interestingly, the [N/Fe] gradient we measure for M87 is starkly different from the [N/Fe] gradients from a sample of comparable galaxies \citep[i.e., the MASSIVE survey;][]{greene2013}. 

However, we also show that the [C/Fe] and [N/Fe] abundances are inconsistent between the galaxy light and the GC stacks. Furthermore, the [C/Fe] and [N/Fe] values for the M87 starlight are correlated rather than anti-correlated, which is a hallmark of the multiple population phenomenon seen in the Local Group GCs \citep[e.g.,][and the references therein]{gratton2004}. Our results therefore show little evidence for stars from dissolved GCs being a significant population in  M87.

The presence of the multiple-population phenomenon has been hinted at by the UV excess in the M87 GCs \citep[][]{peacock2017}. While we find marginal anti-correlation between [C/Fe] and [N/Fe] for the inner halo metal-rich GC stacks, correlation is found in the inner halo metal-poor GC stacks. The metal-poor GCs in the inner halo of M87 were found to be bluer than the Milky Way GC population at fixed metallicity \citep{villaume2019}. That result may be related to differences in horizontal branch morphology between the populations, which, in turn, may be related to light element abundance patterns \citep[e.g.,][and the references therein]{bastian2017}. This, however, is little more than speculation at this point and delving deeper is beyond the scope of this work. 

\subsection{The formation of the outer halo} 
\label{sec:outer_halo}
While observations of GCs and planetary nebulae indicate the presence of an intracluster component that becomes significant beyond $\sim 300$ kpc \citep[][]{longobardi2018a, longobardi2018b}, our work focuses on the material inside this radius and bound to M87. 
The presumption is that the stellar population parameters in the outer halo provide cleaner benchmarks by which to judge accretion predictions since a significant {\it in-situ} population is not expected to complicate the interpretation. 

As discussed in the previous section, the metallicity and [Mg/Fe] gradients in the inner halos of M87 and other massive ETGs are evidence that there was preferential accretion of environmentally quenched satellites. This does not appear to be the case for the outer halo as the outer halo GC stacks have lower [Mg/Fe] abundances than the inner halo stacks. However, when comparing the abundance differences between the inner and outer halo GCs, we need to consider the possibility of mass effects. The outer halo sample consists of more luminous, and therefore more massive, GCs than the inner halo sample (Figure~\ref{fig:cmd}). The abundance spreads in the Milky Way GCs have been shown to correlate with luminosity \citep[Figure 16,][]{carretta2010}, so it would follow that the more massive outer halo GCs might in some way be impacted by this effect. We need to address whether we expect this effect to be significant.

\citet{carretta2010} demonstrated that the correlation between luminosity and extent of abundance spreads (specifically in their case, Na--O) in GCs is driven by the extreme of the abundance anti-correlation, not the median values (see their Figure 11). Since integrated light probes the average parameters of the stellar populations the mass dependency might not be as strong when measuring integrated light. To test this we found the correlation between mass for the Milky Way GCs and [C/Fe], [N/Fe], and [Mg/Fe] as measured by \texttt{alf} from integrated light. For [Mg/Fe] and [C/Fe] we found a mass dependence of $\sim 16\%$. Even accounting for the outer halo GCs being more massive than the most massive GCs, this effect is unlikely to explain the full difference between the inner and outer halo abundances thereby confirming the lower Mg-enhancement in the outer halo compared to the inner halo. 

To determine the specifics of the progenitor satellites that built-up the outer halo in M87, we find that there appear to be at least two distinct populations in the outer halo, in contrast to previous studies which have treated the outer halo GC populations as monolithic \citep[e.g.,][]{forbes2018}. 
First, while the outer halo population has a overall lower metallicity than the inner halo population, the individual GCs display a large range in metallicity. Moreover, even when accounting for the uncertainties in the subpopulation membership assignments when creating the two outer halo stacks, the metallicities of those stacks are different in a statistically significant manner. Second, the metal-poor stack does not display the same unusual $\alpha$-element abundance pattern as the metal-rich stack or the dEs (see Figure \ref{fig:alpha_abundances}).

Even though Mg, Si, and Ca are all $\alpha$ elements, they have different formation sites. Mg is purely a product of massive stars while Si and Ca can both also be produced in Type Ia supernovae 
\citep{woosley2002}. This opens the possibility of using the abundance patterns of the GC stacks to tag the GC population in M87 to possible progenitors. The $\alpha$ element abundance pattern displayed by the metal-rich stack is echoed by the dE population in the Virgo cluster (Figure~\ref{fig:alpha_abundances}). This makes it tempting to point to the dE population as the progenitors of some portion of the outer halo of M87. However, it is important to acknowledge the difficulty in abundance tagging in this scenario because of the strong radial gradients in dEs \citep[Figure 1;][]{sybilska2018}. \citet{sen2018} used the $R_e/8$ aperture with the nucleus included, while \citet{sybilska2018} took the luminosity-weighted average of spectra within $1R_e$.
While noting this caveat, the metallicity gradient for the outer halo metal-rich subpopulation is negative and inconsistent with a flat gradient within the $1\sigma$ uncertainty, which is consistent with the bulk of this population being brought in by low-mass satellites.

Furthermore, the progenitor mass implied by the median metallicity of the outer halo metal-rich GC population is $M_* \sim 10^9 M_\odot$ \citep[][]{gallazzi2005} which is consistent with the predictions from IllustrisTNG. At $>100$ kpc, \citet{pillepich2018} predicted that 90\% of the {\it ex-situ} mass comes from progenitors with stellar masses $\gtrsim 5\times10^9 M_\odot$, with the typical progenitor mass being $\sim 7 \times 10^{10} M_\odot$ (see their Figure 13b). However, as in the inner halo, the metal-poor population is not consistent with the predictions from simulations.

Other aspects of the GC system are consistent with our interpretation of the stellar population parameters.
In the outer halo, M87 has a $V$-band luminosity of $\sim 2.9\times10^{10} L_\odot$, very similar to the total luminosity of the Milky Way \citep{kormendy2009, bhawthorn2016}. From S\'ersic fits to the photometric sample of M87 GCs from \citet{strader2011b} and correcting for GC luminosity function incompleteness, we estimate there are $\sim1200$ metal-rich GCs and $\sim 4500$ metal-poor GCs in this region. The GC specific frequency ($S_N$) of the outer halo is then $S_N\sim16$; which besides M87, the only galaxies in Virgo with such a high value are dwarfs with $M_V\sim -17$ to $-16$ and fainter \citep[$M_* \sim 10^8 \-- 10^9 M_\odot$,][Figure 12]{peng2008}. 

A similar conclusion to our own was reached by \citet{longobardi2018b} using the outer halo light color M87 to infer low-mass progenitors. \citet{hartke2018} took that result to indicate a problem with the feedback prescription in IllustrisTNG. However, we also need to consider the possibility that the accreted satellites were different than the surviving population. That is, that there {\it may} have been Milky Way-mass galaxies but with $S_N$ more like dwarfs, that no longer exist today. Or possibly similar to the ultra-diffuse galaxies that have been found to have very high $S_N$ \citep{peng2016}.
The evidence for a dynamically cold phase-space shell prompted \citet{romanowsky2012} to suggest that $\sim 20\%$ of M87's GC population within $50 < R_{\rm gal}~{\rm kpc} < 95$ came from an E/S0 progenitor with luminosity $\sim 0.5L^*$, which could lead to flattened the metallicity gradient we measure in the outer halo metal-poor GC subpopulation.

\section{Summary}

Using updated full-spectrum SPS models we present the first detailed stellar population analysis of M87 and its GC system from spectroscopy. We applied the models to 322 GCs extending from the inner to outer halo. We use these same models to fit IFU spectroscopy to get spatially-resolved stellar population parameters of M87 itself.

We present a new statistical framework to measure the radial metallicity gradients of a multimodal GC system that accounts for the covariance between subpopulation membership assignments and the physical parameters of the subpopulations while doing a statistical de-projection of the galactocentric distances which enables much more accurate measurement of the linear gradient parameters of the GC subpopulations. Our main results are as follows:

\begin{itemize}
\item We show the first direct spectroscopic comparison of field stars and GCs in M87, confirming the association of field stars and red GCs. 
\item In the inner halo, we measure for both the metal-rich and metal-poor subpopulations remarkably flat metallicity gradients. With the additional constraint of the galaxy stellar metallicity gradient, we find this to be compelling circumstantial evidence for a population of metal-poor GCs that formed {\it in-situ} directly in the halo. The expectation for the alternative scenario, that the low-mass satellites preferentially came in early and were in place by the time the major mergers began, is something that should be checked in cosmological simulations of galaxy evolution.
\item From the $\alpha$ abundances of the outer halo GC stacks, we find evidence for relatively recent accretion of low-mass satellites with extended star-formation histories, unlike the the inner halo which shows evidence for accretion of early quenched satellites.
\item We find evidence for  a metal-poor, $\alpha$-enhanced population in both the inner and outer halo, not anticipated by current cosmological galaxy simulations. It currently unclear whether this is the sign of persistent problems in the subgrid physics of the cosmological simulations, or indication of a missing satellite population around massive ETGs.
\end{itemize}

\acknowledgments

We would like to thank Karl Gebhardt and Sarah Bird for generously sharing data with us and to E. Cunningham, N. Mandelker, and S. Woosley for discussions on various aspects of this paper. We greatly appreciate the anonymous referee's thoughtful report that helped us improve the quality of this manuscript. AV would like to acknowledge the NSF Graduate Fellowship Program for its support. JS was supported by NSF Grant AST-1514763 and the Packard Foundation. AJR was supported by National Science Foundation grant AST-1616710, and as a Research Corporation for Science Advancement Cottrell Scholar. The authors would like to thank the Center for Computational Astrophysics at the Flatiron Institute for enabling this collaboration.

\software{IPython \citep{PER-GRA:2007}, 
          SciPy \citep{2020SciPy-NMeth},
          NumPy  \citep{van2011numpy}, 
          matplotlib  \citep{Hunter:2007},
          Astropy \citet{astropy18},
          PyMC3 \citep{pymc3}, 
          emcee \citep{fm2013}
}

\bibliography{references}{}
\bibliographystyle{aasjournal}

\appendix 

\setcounter{table}{0}
\renewcommand{\thetable}{\Alph{section}\arabic{table}}

\section{Derivation of Likelihood Function for True Distances}
\label{ap:derivation}

While the true 3D distance is given by $|\vec{r}| = \sqrt{x^2 + y^2 + z^2}$, what we actually observe is $r_\perp = \sqrt{x^2 + y^2}$. The $xy$ coordinates can be written in terms of $r_\perp$, 

\begin{align*} 
    x = & r_\perp cos\theta \\
    y = & r_\perp sin\theta\text{, and then,} \\
    z = & r_\parallel \quad.
\end{align*}

Our model defines the distribution in $x$, $y$, and $z$ as Gaussian, but it is useful to, instead, reparameterize in terms of $r_\perp$ and $r_\parallel$.
In order to maintain the same density through this change of variables, we need to take the Jacobian of the transformation into account.
Specifically,
\begin{eqnarray}
\left | p(x,\,y,\,z)\,\mathrm{d}x\,\mathrm{d}y\,\mathrm{d}z \right| &=& \left | p(r_\perp,\,\theta,\,r_\parallel)\,\mathrm{d}r_\perp\,\mathrm{d}\theta\,\mathrm{d}r_\parallel \right|\\
p(r_\perp,\,\theta,\,r_\parallel) &=& |J|\, p(x,\,y,\,z)
\end{eqnarray}
where $p(x,\,y,\,z)$ is Gaussian and $|J|$ is the absolute value of the determinant of the Jacobian matrix
\begin{eqnarray}
|J| &=& \left|\begin{matrix}
\frac{\mathrm{d}x}{\mathrm{d}r_\perp} & \frac{\mathrm{d}y}{\mathrm{d}r_\perp} & \frac{\mathrm{d}z}{\mathrm{d}r_\perp}\\
\frac{\mathrm{d}x}{\mathrm{d}\theta} & \frac{\mathrm{d}y}{\mathrm{d}\theta} & \frac{\mathrm{d}z}{\mathrm{d}\theta}\\
\frac{\mathrm{d}x}{\mathrm{d}r_\parallel} & \frac{\mathrm{d}y}{\mathrm{d}r_\parallel} & \frac{\mathrm{d}z}{\mathrm{d}r_\parallel}\\
\end{matrix}\right| \\
&=& \left|\begin{matrix}
\cos\theta\quad & \quad\sin\theta\quad & \quad 0 \\
-r_\perp\sin\theta\quad & \quad r_\perp\cos\theta\quad & \quad 0\\
 0 \quad & \quad 0 \quad & \quad 1\\
\end{matrix}\right| \\
&=& \left|r_\perp\,(\sin^2\theta + \cos^2\theta)\right| = r_\perp \quad.
\end{eqnarray}
Therefore,
\begin{eqnarray}
p(r_\perp,\,\theta,\,r_\parallel) &=& \frac{r_\perp}{(2\,\pi\,R^2)^{3/2}}\exp\left(-\frac{x^2 + y^2 + z^2}{2\,R^2}\right) \\
&=& \frac{r_\perp}{(2\,\pi\,R^2)^{3/2}}\exp\left(-\frac{r_\perp^2 + r_\parallel^2}{2\,R^2}\right)\quad.
\end{eqnarray}
Finally, under our assumption of isotropy, we can marginalize over the position angle $\theta$ to find
\begin{eqnarray}
p(r_\perp,\,r_\parallel) &=& \int p(r_\perp,\,\theta,\,r_\parallel)\,\mathrm{d}\theta \\
&=& \frac{r_\perp}{\sqrt{2\,\pi}\,R^3}\exp\left(-\frac{r_\perp^2 + r_\parallel^2}{2\,R^2}\right) \\
&=& \mathrm{Rayleigh}(r_\perp;\,R)\,\mathrm{Normal}(r_\parallel;\,0,\,R)\quad.
\end{eqnarray}

\section{Individual References for JINABase data}
\label{ap:jinarefs}

\begin{deluxetable}{lcccc}
\tabletypesize{\footnotesize}
\tablecolumns{8}
\tablewidth{0.9\columnwidth}
\tablecaption{Individual references for JINAbase compilation of dwarf galaxy field stars used in this paper. \label{table:jinaref_dwarfs}}
\tablehead{
 \colhead{Reference}  & \colhead{[Fe/H]} & \colhead{[Mg/Fe]} & \colhead{[Si/Fe]} & \colhead{[Ca/Fe]}
}
\startdata
\centering
\citet{COH09}  &   x & x & x& x\\
\citet{FEL09}  &   x & x & --& --\\
\citet{FRA16}  &   x & x & --& --\\
\citet{FRE10a} &   x & x & x& x\\
\citet{GEI05}  &   x & x & x& x\\
\citet{GIL13}  &   x & x & x& x\\
\citet{ISH14}  &   x & x & --& --\\
\citet{JI16b}  &   x & x & x& x\\
\citet{KOC08}  &   x & x & x& x\\
\citet{SHE01}  &  x & x & x& x\\
\citet{SHE03}  &  x  & x & x& x\\
\citet{SKU15}  &   x & x & x & x\\
\enddata
\end{deluxetable}

\begin{deluxetable}{lcccc}
\tabletypesize{\footnotesize}
\tablecolumns{8}
\tablewidth{0.9\columnwidth}
\tablecaption{Individual references for JINAbase compilation of Milky Way field stars used in this paper \label{table:jinaref_mw}}
\tablehead{
 \colhead{Reference}  & \colhead{[Fe/H]} & \colhead{[Mg/Fe]} & \colhead{[Si/Fe]} & \colhead{[Ca/Fe]}
}
\startdata
\centering
\citet{ALL12}  &    x & x & -- & --\\
\citet{AOK02a} &    x & x & --  &-- \\
\citet{AOK02d} &    x & x & x & x \\
\citet{AOK05}  &   x & x & x & x \\
\citet{AOK07a} &   x & x & --  &-- \\
\citet{AOK08}  &   x & x & --  &-- \\
\citet{AOK12}  &    x & x & --  &-- \\
\citet{AOK13}  &    x & x & --  &-- \\
\citet{AOK14}  &    x & x & --  &-- \\
\citet{BAB05}  &   x & x & --  &-- \\
\citet{BAR05}  &   x & x & --  &-- \\
\citet{BEN11}  &    x & x & x & x\\\
\citet{CAR02}  &   x & x & x & x\\
\citet{CAY04}  &    x & x & x & x\\
\citet{COH03}  &   x & x & x & x\\
\citet{COH04}  &    x & x & x & x\\
\citet{COH06}  &    x & x & x & x\\
\citet{COH13}  &   x & x & x & x\\
\citet{COW02}  &   x & x & x & x\\
\citet{CUI13}  &   x & x & x & x \\
\citet{FUL00}  &  x & x & x & x\\
\citet{HAN15}  &    x & x & --  & -- \\
\citet{HOL15}  &   x & x & --  &-- \\
\citet{HON04} &    x & x & x & x\\
\citet{HOW15}  &   x & x & x & x\\
\citet{HOW16}  &   x & x & x & x \\
\citet{ISH10}  &   x & x & x & x\\
\citet{IVA03}  &    x & x & x & x\\
\citet{IVA06}  &    x & x & x & x \\
\citet{JAC15}  &   x & x & x & x\\
\citet{JOH02a} &   x & x & x &x \\
\citet{JOH04}  &    x & x & --  &-- \\
\citet{JON05}  &   x & x & x & x \\
\citet{JON06}  &    x & x & --  & --  \\
\citet{KOC15}  &  x & x & x & x\\
\citet{LAI08}  &    x & x & x &x \\
\citet{LI15a}  &    x & x & x & x\\
\citet{MAS12}  &    x & x & --  & -- \\
\citet{MCW95}  &    x & x & x &x\\
\citet{NOR97a} &   x & x & x &x \\
\citet{PLA15b} &    x & x & x &x\\
\citet{PRE00}  &   x & x & --  &-- \\
\citet{PRE01}  &    x & x & --  &-- \\
\citet{PRE06}  &   x & x & x &x\\
\citet{ROE08}  &    x & x & --  &-- \\
\citet{ROE10}  &    x & x & x &x\\
\citet{ROE14b} &   x & x & x &x\\
\citet{RYA91}  &   x & x & x &x\\
\citet{SIQ14}  &    x & x & x &x\\
\citet{ZAC98}  &    x & x & --  &-- \\
\citet{ZHA09}  &    x & x & x &x\\
\enddata
\end{deluxetable}

\end{document}